\documentclass[referee]{aa}
\newcommand{\apropto}{\;
  \raise0.3ex\hbox{$\propto$\kern-0.75em\raise-1.1ex\hbox{$\sim$
  }}\;\hskip-2pt }
\usepackage[dvips]{graphicx}
\usepackage[usenames]{color}
\usepackage[T1]{fontenc}
\newcommand{\lta}{\;
  \raise0.3ex\hbox{$<$\kern-0.75em\raise-1.1ex\hbox{$\sim$
  }}\;\hskip-2pt }
\newcommand{\gta}{\;
  \raise0.3ex\hbox{$>$\kern-0.75em\raise-1.1ex\hbox{$\sim$
  }}\;\hskip-2pt }

\begin{document}
\title{Magnetic fields in ring galaxies
}

   \author{D.Moss\inst{1}
  \and   E. Mikhailov\inst{2}
   \and  O. Silchenko\inst{3}
        \and D. Sokoloff\inst{2}
          \and C. Horellou\inst{4}
          \and R. Beck\inst{5}}

   \offprints{D.~Moss}

   \institute{ School of Mathematics, University of Manchester,
               Oxford Road, Manchester, M13 9PL, UK
               \and
               Department of Physics, Moscow University, 119992 Moscow, Russia
               \and  Sternberg Astronomical Institute and Department of Physics, Moscow M.V. Lomonosov State University, Universitetskij 
               pr., 13, 119992 Moscow, Russia               
               \and Chalmers University of Technology, Dept. of Earth and Space Sciences, Onsala Space 
               Observatory, SE-439 92 Onsala, Sweden
               \and MPI f\"ur Radioastronomie, Auf dem H\"ugel 69, 53121 Bonn, Germany}

   \date{Received ..... ; accepted .....}

\abstract{Many galaxies contain magnetic fields supported by galactic dynamo action. The investigation of  these
magnetic fields can be helpful for understanding galactic evolution. However,
nothing definitive is known about magnetic fields in ring galaxies.}
{Here we investigate large-scale magnetic fields in a previously unexplored context,
namely ring galaxies, and 
concentrate our efforts on the structures that appear most promising for galactic dynamo action, i.e. outer star-forming rings in 
visually unbarred galaxies.}
{We use tested methods for modelling $\alpha-\Omega$ galactic dynamos, taking into account the available observational 
information concerning ionized interstellar matter in ring galaxies.}
{Our main result is that dynamo drivers in ring galaxies are strong enough to excite large-scale magnetic fields
in the ring galaxies studied.  The variety of dynamo driven magnetic configurations in ring galaxies 
obtained in our modelling is much richer than that found in
 classical spiral galaxies. In particular, 
various long-lived transients are possible. An especially interesting case is 
that of NGC~4513 where the ring 
counter-rotates with respect to the disc. Strong shear in the region between the disc and the ring is
associated with unusually strong dynamo drivers for the counter-rotators. The effect of the strong drivers is found to be unexpectedly moderate.
With counter-rotation in the disc,
a generic model shows that a steady mixed parity magnetic configuration, 
unknown for classical spiral galaxies,
may be excited, although we do not specifically model NGC~4513.}
{We deduce that ring galaxies constitute a morphological class of galaxies 
in which identification of large-scale magnetic fields 
from observations of polarized radio emission, as well as dynamo modelling,
may be possible. Such studies 
have the potential to throw
additional light on the physical nature of rings, their lifetimes and evolution.
}


\titlerunning{Magnetic fields in ring galaxies}
\authorrunning{D.~Moss et al.}

\maketitle

\section{Introduction}
\label{int}

The interstellar medium of spiral galaxies contains large-scale
magnetic fields (of scale comparable with the galactic size)
which are believed to be excited by a 
galactic dynamo, that is driven by the joint action of differential rotation
and mirror-asymmetric turbulent flows. As we
know from observations of several nearby spiral galaxies such as M~51,
as well as the Milky Way, this large-scale magnetic field is mainly azimuthal
in the disc,
with field strength of several $\mu$G, i.e. the magnetic energy is
close to equipartition with the kinetic energy of turbulent motions, see
e.g. Beck (2016).
In addition to the large-scale magnetic field, a small-scale component of comparable strength is present in the interstellar medium.
At least in  the large isolated spiral galaxy NGC~6946, the large-scale magnetic field
is organized in magnetic arms located between the stellar arms. These features are more or less related to predictions of galactic dynamo theory 
(for a review see e.g. \cite{Beck96}). There are some attempts (e.g. \cite{Arshakian09}) to follow magnetic field
evolution in the context of galactic evolution.

\begin{figure*}
\vspace*{2.5cm}
\resizebox{\hsize}{!}
{\includegraphics{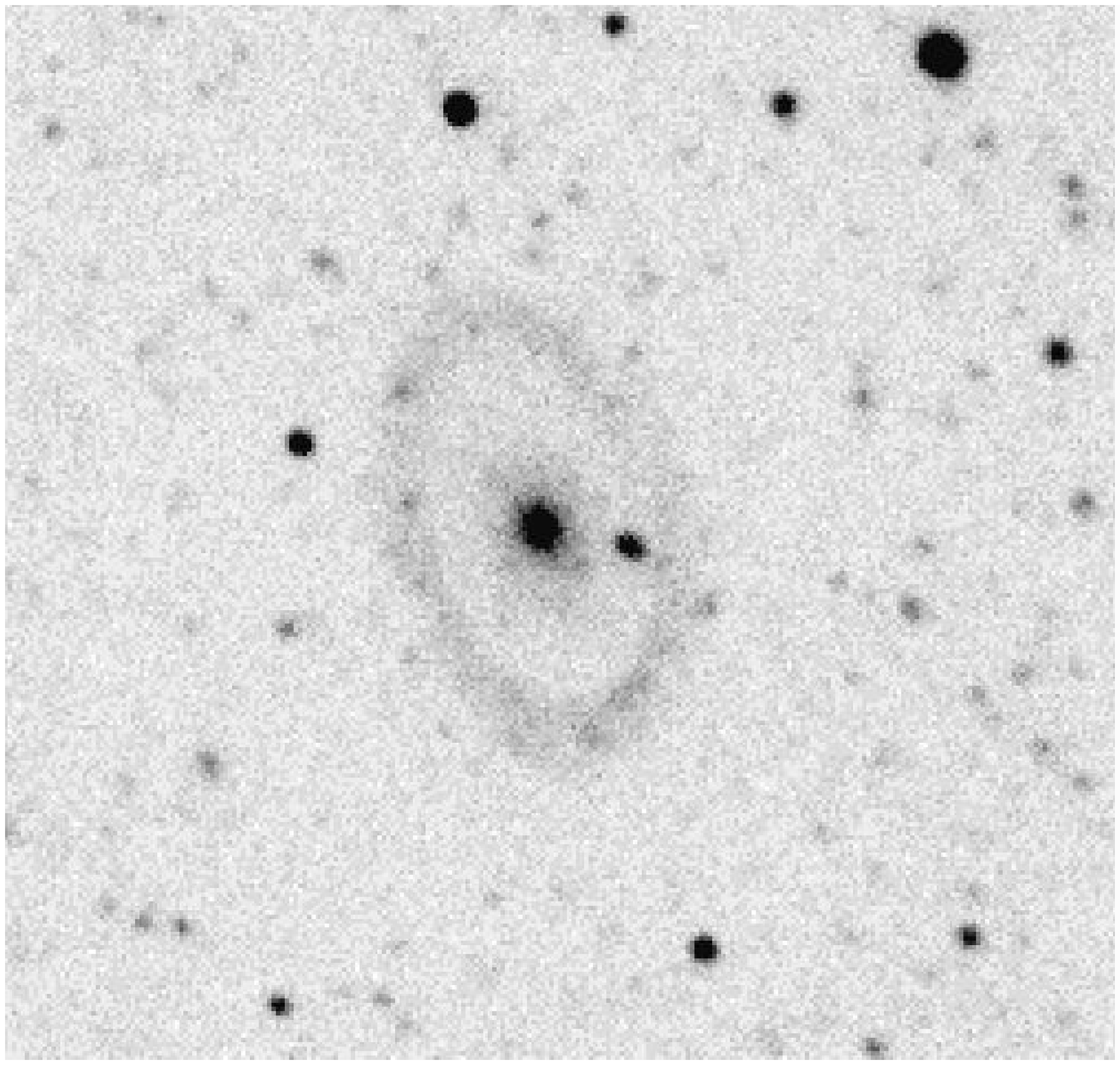}
\includegraphics{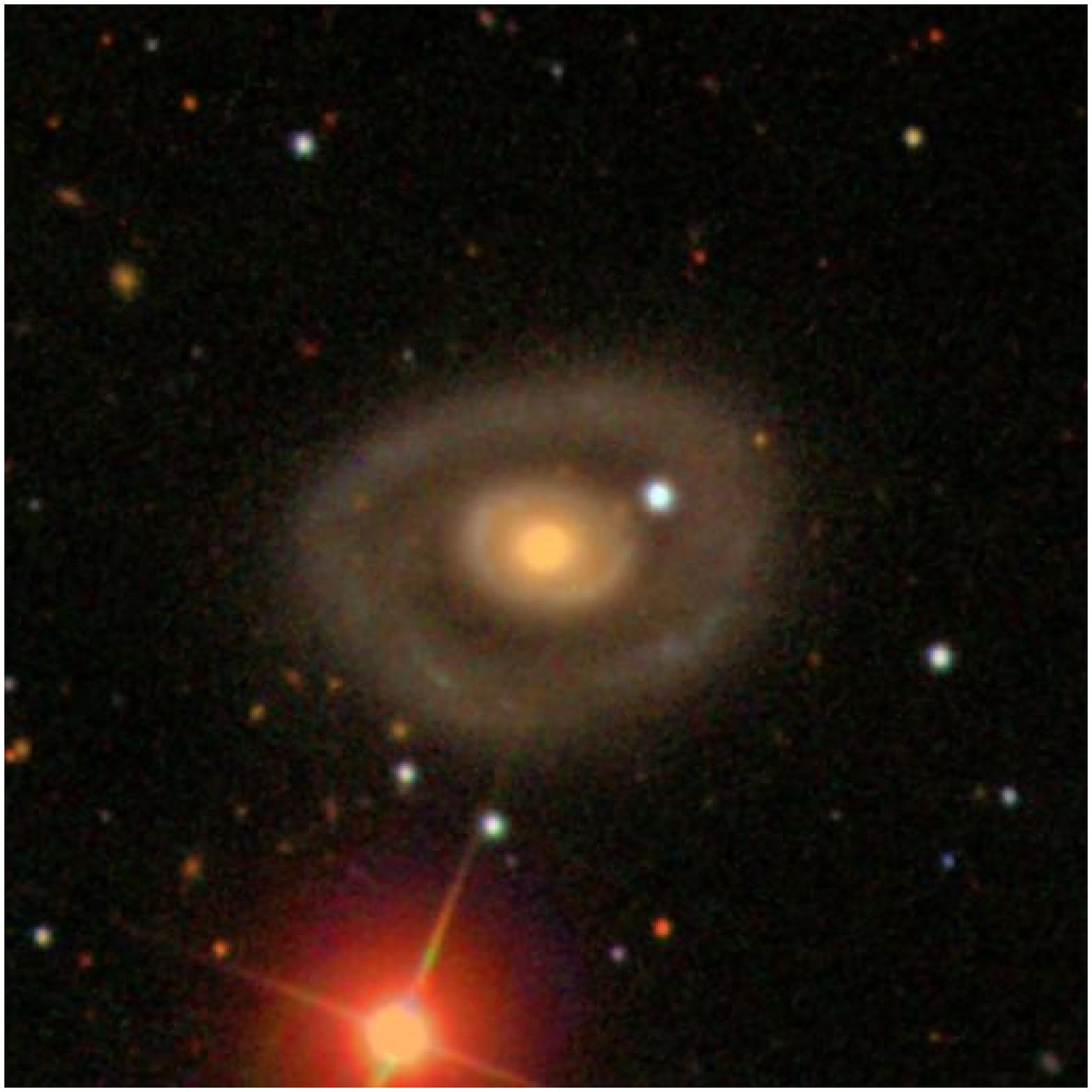}
\includegraphics{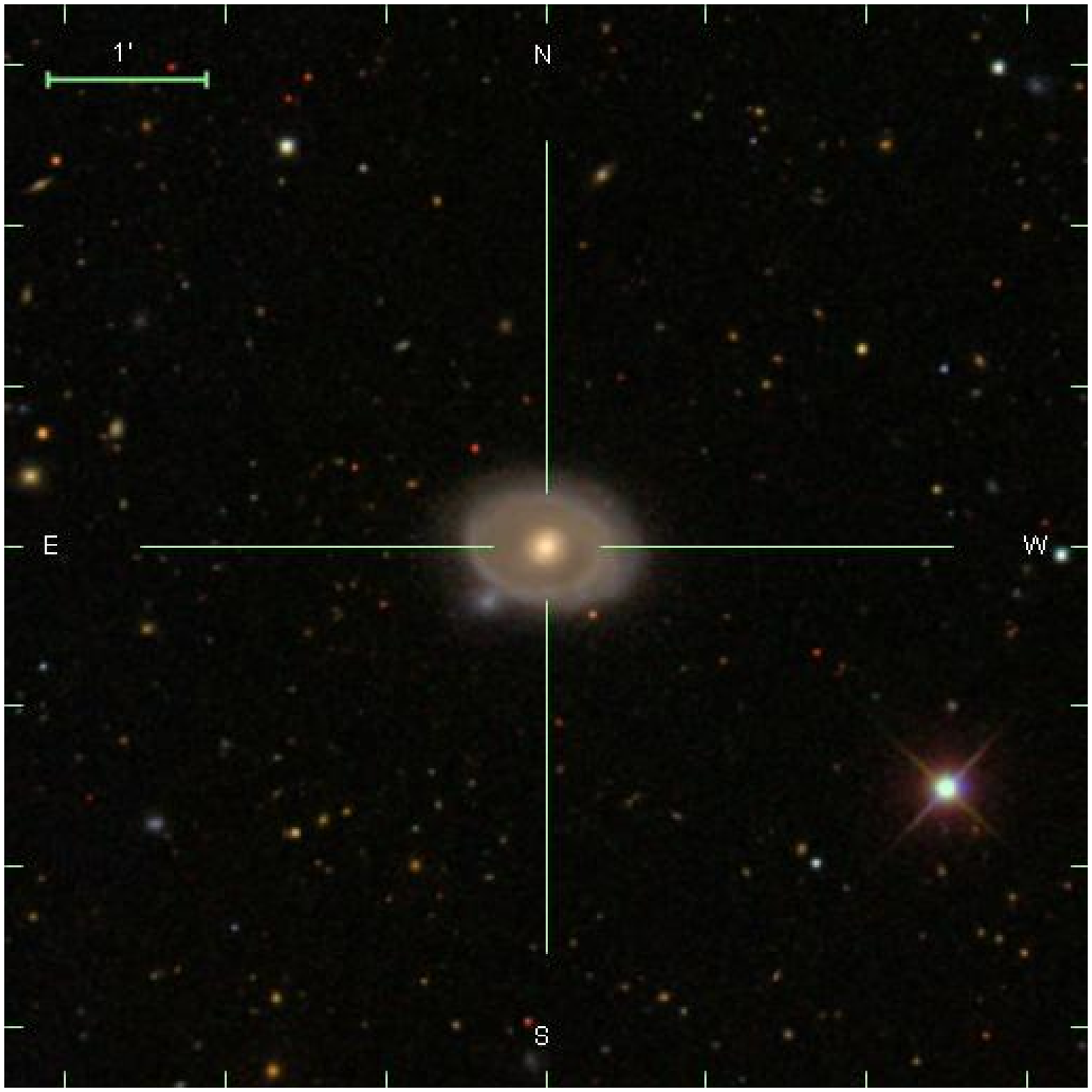}
\includegraphics{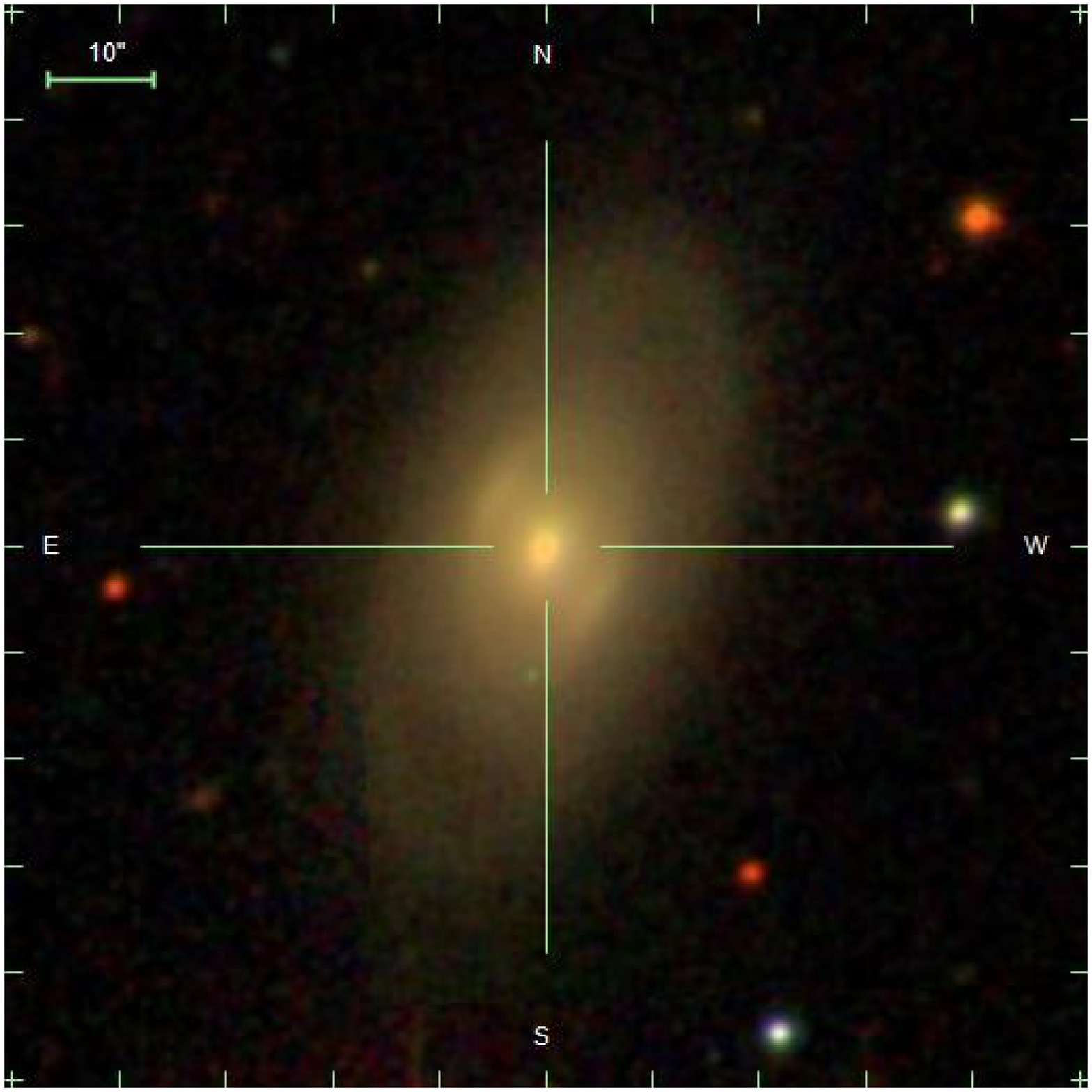}}
\hspace*{1cm}
{ \Large   NGC~4513 $\mbox{ }$ \quad \quad \quad $\mbox{ }$  IC~5285  \quad \quad \quad  $\mbox{ }$ \quad \quad \quad UGC~5936  $\mbox{ }$  \quad \quad \quad
UGC~9980 $\mbox{         }$} 
\caption{Examples of 
lenticular (they have no spiral arms but do have discs)
galaxies with outer rings: composite-coloured images from the SDSS survey and image from ultraviolet GALEX survey for NGC~4513. (The NGC~4513 image from
the SDSS survey shows the ring with a rather low contrast.)} \vspace*{2cm}
\end{figure*}

Observations  and modelling of magnetic fields of barred galaxies,
considered as a particular
morphological class of galaxies, have been combined with interpretation of 
the results obtained in the framework of galactic dynamos
(\cite{Metal98,Betal99,Metal01}). Previous experience with barred galaxies has shown 
the investigation
of magnetic fields in galaxies of particular morphological classes to be fruitful (\cite{Betal02,Betal05,Metal07}), e.g. specific magnetic
structures near bars were identified and interpreted as being important in feeding
black holes at galactic centres.

Further studies of investigation of magnetic fields in particular morphological classes of galaxies should
include examples of classes which appear to be candidates  for
future observations of polarized radio emission, which
is an important indicator of the presence of large-scale magnetic fields.
The aim of this paper is to discuss ring galaxies in this context;
note however that radio polarization detections of these galaxies are scarce.

Ring galaxies were defined as a class of disc galaxies by compiling
a list and atlas of galaxies in which the rings were separated from the bright central parts of the galaxies (\cite{vv60}). He stated
that the rings were "structural elements on the same level as bars, spiral arms, and discs". The typical
radii of the outer galactic rings range between 5 and 25 kpc (\cite{theys_spi}), and the width might be as large as a few
kpc. Though the rings are certainly stellar structures, they often contain gas and demonstrate some level of star formation.
For example, a recent catalogue of ring structures derived from The Spitzer Survey of Stellar Structure in Galaxies
(\cite{s4g}),
ARRAKIS (\cite{arrakis}), includes outer stellar rings detected in  50\%--60\%\ of all nearby S0 galaxies (\cite{lauri11,arrakis}); and among those,
half of all rings in S0s are seen in the ultraviolet -- in the NUV-band being mapped by the GALEX space
telescope -- so revealing recent star formation (\cite{kostuksil}). In early-type (S0) disc galaxies,
the star formation is mainly organized in large-scale rings (\cite{pogge_esk,salim}), and the neutral-hydrogen distributions 
are also ring-like (\cite{vandriel}). The gas in the outer ultraviolet rings is often ionized by young stars (\cite{Silchenko14}).
van Driel \& van Woerden (1991), who had traced the rotation of the S0 discs far beyond the centre by observing neutral hydrogen in the rings,
noted that the rotation curves in the outer discs were flat and so the rotation was obviously differential.
In other words, the interstellar matter in ring galaxies has all the features that drive dynamo action in classical nearby
galaxies with magnetic field. The natural question here is: are the dynamo drivers in ring galaxies strong
enough to get self-excitation of magnetic fields? 
Continuum radio emission, indicative of the presence of magnetic fields,
has been detected from a number of ring galaxies, e.g. Ghigo (1980), Jeske(1986), Appleton et al. (1999);
and there is a possible detection of polarized emission from
the polar ring galaxy NGC~660
(Wiegert et al 2015).
Estimates presented below
are optimistic enough to suggest that  further
observational studies of these systems may be justified.

\begin{figure}
\begin{center}
\includegraphics[width=9cm]{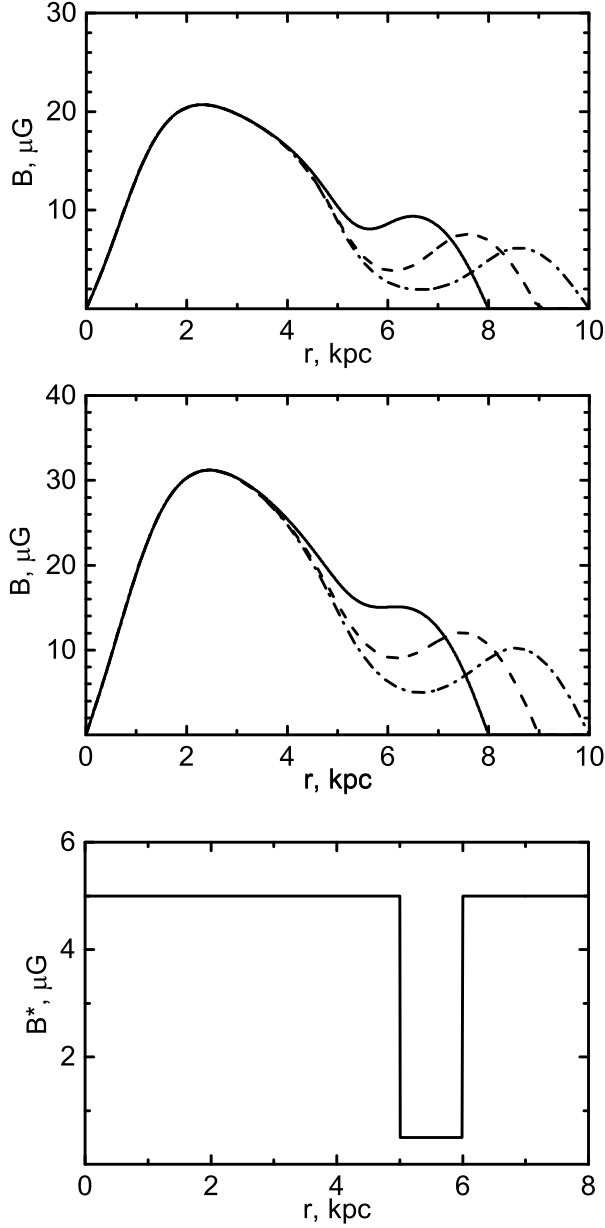}
\end{center}
\caption{Regular magnetic field generation in the ring for various  
$d$ ($2d$ is the width of ring) and dynamo numbers $D$. The upper panel shows the magnetic field for $t=10^{10} \mbox{ yr}$ for $D=20,$ the 
middle one the same for $D=50.$ The solid curves are for a ring with $d=1 \mbox{ kpc},$ the dashed curve,  $d=1.5 \mbox{ kpc},$ and the dot-dashed curve is for  $d=2 \mbox{ kpc}$. The lower panel shows the equipartition field 
$B^*$ (which is determined by the kinetic energy of the turbulent gas) 
versus radius for a representative case with $L=5$ kpc, $d=1$ kpc.  }
\label{fig1}
\end{figure}

In the framework of our paper we have to recognize that the physical properties of the interstellar medium in
ring galaxies, and also the dynamo context of the problem,
are much less investigated than the corresponding topics for such galaxies
as the Milky Way or several nearby galaxies  which are considered to be the site of typical examples of galactic dynamos  (but see, e.g.
\cite{Cathy95}).
Their properties are quite diverse. Correspondingly, we are interested here in crude preliminary estimates rather than in detailed modelling. 
Moreover, among the ring galaxies of quite diverse origins (see the next Section) we can select the cases which
are particularly promising for the dynamo development. Fortunately, rings are often found in the same plane as 
the main
stellar disc component\footnote{There are examples of polar rings, i.e. rings inclined to the disc plane 
(\cite{polarrings}).  
However it is believed that after being accreted
with arbitrary inclined angular momentum, gaseous rings settle to the plane of the galactic disc in a few orbital 
times, see
e.g. Steiman-Cameron \& Durisen (1988), Christodoulu \& Tohline (1993), Colley \& Sparke (1996).} This means that we can consider ring galaxies as
mainly flat systems and use various simplifications of the general dynamo equations previously developed for 
spiral galaxies.
Additionally, the axisymmetry of  the discs of  ring galaxies simplifies dynamo studies.

\section{Interstellar gas in ring galaxies}

There are three distinct mechanisms that have been proposed for the
origin of the marked outer ring structures in disc galaxies. 
The most exotic, and so in practice probably the rarest, is
the impact mechanism (\cite{freemanvauc,LyndTo76,theys_spi,fewmadore,AppStr,atha97}).
This mechanism involves the infall of a dwarf satellite or of a high-velocity massive 
cloud impacting very close to the galactic
centre; this suggests that the initial configuration will be a strongly inclined, almost radial orbit of the satellite 
or of the cloud around the host galaxy. Such collisions do not destroy the large-scale stellar disc 
of the host galaxy but instead can provoke a compression wave running outward through the large-scale 
gaseous (and stellar) disc of the host galaxy.  The shock compresses the gas of the galaxy, initially  
quasi-homogeneously
distributed over the whole disc, and 
the compression becomes sufficient to initiate 
star formation in a ring.  Large spiral galaxies with high gas content appear to be the most favourable for this mechanism to operate.

\begin{figure}
\begin{center}
\includegraphics[width=9cm]{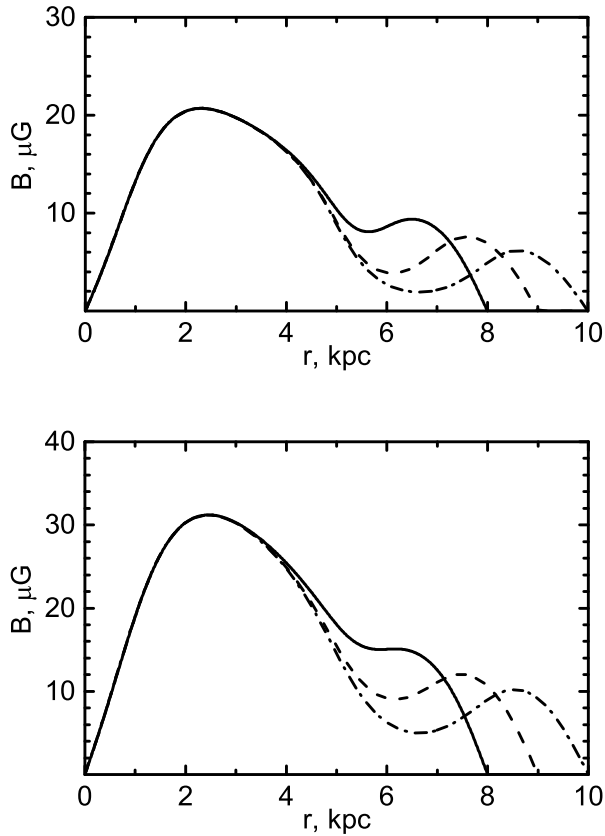}
\end{center}
\caption{Dynamo generated magnetic fields for various values of $D$ and $L$ when the seed field is concentrated in the inner disc only. The ring half-width $d=1$ kpc in all cases.  The upper panel
shows the magnetic field for $t=10^{10} \mbox{ yr}$ when the dynamo number
 $D=20,$ the lower shows the same for $D=50.$ The solid curves
are for the gap extent $L=1 \mbox{ kpc},$ the dashed curves  for $L=2 \mbox{\, kpc},$ and the dot-dashed curves are 
for $L=3\mbox{\, kpc}.$}
\label{fig2}
\end{figure}

Another scenario,  perhaps more frequent, invokes resonance (\cite{schommer_sullivan, atha82,buta95,butacombes}).
The resonance scenario requires violation of axisymmetry of the galactic disc and can occur in the 
presence of a bar or at least of a slightly oval bulge (\cite{jung}). The triaxial gravitational 
potential causes radial gas motions so that the gas can be acquired  at the inner and outer Lindblad resonances 
of the non-axisymmetric rigidly rotating pattern.

The third possibility for ring formation involves outer gas accretion (\cite{butacombes}). 
We believe that
this scenario is probably the most common among early-type discs -- namely, among lenticular galaxies --  which possess outer rings 
in more than a half of all cases (\cite{arrakis}). Interestingly, the frequency of strong bars diminishes among
lenticular galaxies compared to spirals (\cite{lauri11,lauri13}), while the frequency of outer rings rises 
(\cite{arrakis}). 

This implies a rather low probability that the outermost rings in lenticular galaxies are 
due to resonance.
The high fraction of observed gas counter-rotation with respect to the stellar components, especially among 
the S0s in rarified environments, strongly supports an origin through accretion (\cite{Katkov15}). Some examples 
of the
outer star forming  rings in visually unbarred 
galaxies are shown in Fig.~1. This figure presents a particular class of ring galaxies -- the accreted 
star forming
rings in axisymmetric S0s which are especially suitable to host a dynamo process. 
The class is however far from being homogeneous. Fig.~1 illustrates that the ring galaxies of interest 
can vary considerably in appearance. 
This is why we
have to consider a variety 
of possible dynamo effects which might contribute to magnetic field formation in such 
scenarios and experiment with the dynamo numbers that estimate
the strengths of the effects. A particular challenge is that NGC 4513, 
shown in the first panel of Fig.~1, has a counter-rotating ring.
IC 5285, UGC 5936 and UGC 9980 do not demonstrate 
observable radial gas velocities, thus ruling out a collisional origin 
for their rings 
(\cite{uvrings,Silchenko14,Katkov15}).

In fact, to discriminate between the origin scenarios for an outer ring in a particular galaxy, kinematical
data are necessary, and 3D (two spatial coordinates on the plane of sky and one
along the wavelength axis, \cite{C95}) spectral data are especially useful (\cite{moisrev}). The collisional 
rings and 
OLR (Outer Lindblad Resonance) rings generate radial gas outflow while accreted gaseous rings rotate circularly,
though sometimes outside the main galactic plane. In addition, the optical spectral data provide the possibility
of determining the excitation mechanism for the ionized (warm) gas: the flux ratios of strong emission lines
discriminate easily between the mechanisms of shock excitation and of ionization by UV radiation of young stars 
(\cite{bpt,vo87}).
To facilitate the work of the dynamo in magnetic-field generation, we need
\begin{itemize}
\item{long-lived (order of Gyrs) rings; so collisional rings, with their lifetimes of a few hundred Myr, are unlikely to be candidates;
}
\item {a supply of free electrons, so 
an ionized state of the gas which can be provided and supported
by continuous star formation in the rings;
}
\item{regular circular gas rotation in the main galactic planes, presumably with a flat rotation curve, to
provide differential character of the gas rotation.
}
\end{itemize}

In some ways rings generated by accretion maybe be the most favourable
for our investigation, with estimated lifetimes possibly approaching
the Hubble time (Buta \& Combes 1996).
Based on the results of \cite{uvrings,Silchenko14,Katkov15} we
select a small sample with the properties mentioned above and present it in Fig.~1. The typical properties
of these galaxies are to be used in the further theoretical consideration of the dynamo magnetic-field generation 
in ring galaxies. Some extension of the typical picture can be also considered, 
by taking into account 
the possible cases of the outer gas counter-rotation, as in NGC~4513 (\cite{Silchenko14}), or strong flaring
of the gas layer in the ring area, as in UGC~5936 (\cite{uvrings}).
Of this sample, continuum radio emission has been detected from IC~5285 (\cite{Co98}).

\section{Dynamos in ring galaxies}

From the viewpoint of galactic dynamos the main specific problem of ring galaxies 
is to what extent 
drivers of the galactic dynamo, i.e. differential rotation and mirror-asymmetric turbulence,
can produce a magnetic structure in the ring that is independent of that in the disc, 
or is propagation of magnetic 
field generated in the disc into the  ring a more realistic process? Of course, the answer depends on properties 
of the interstellar medium near the boundary between disc and ring. One possibility here is that rings are 
overdense  regions of warm gas superposed on 
an exponential disk as in late-type disc galaxies. The other possibility is that there is a gap in density 
distribution between disc 
and ring.  The latter option which is the case of lenticular galaxies where 
warm ionized gas is often observed only in the rings (\cite{pogge_esk,salim}) is obviously favourable for formation 
of independent magnetic structure in the ring and looks realistic especially for counter-rotating rings.    

In the rings, the dynamo
drivers have in addition to overcome the turbulent diffusion of the large-scale magnetic field into the gap 
between
the disc and the ring. We recall that in the standard disc dynamo  the 
drivers only have to compensate losses of
the large-scale magnetic field mainly through the upper and lower boundaries of the disc.
We start with an 
order of magnitude estimate of the role of these losses.

We need to compare somehow the results obtained for ring galaxies with 
those for spiral galaxies. We 
proceed as follows. The conventional estimate for the intensity of dynamo action (e.g. \cite{Betal05}) is given in terms 
of the dynamo number $D = 9 (\Omega h/v)^2$ where $\Omega$ is angular velocity, $h$ is disc thickness 
and $v$ the rms turbulent velocity. 
Usually $\Omega h \approx v$ and $D \approx 10$. This estimate is based on 
the assumption that $r d \, \Omega/ d \, r = \Omega$,
which is an obvious idealisation for a flat (or outer Brandt) 
rotation curve. For real rotation curves it seems reasonable to try 
values $D$ of several dozen, i.e. significantly larger than the normal
estimates. This logic is directly 
applicable for rings that co-rotate with the disc.  
However a galaxy with a counter-rotating ring has unusually 
strong shear in the region between the disc and ring. 
We  attempt to parametrize this situation as follows. We present the 
rotation curve as a combination of two  disjoint pieces of a Brandt 
rotation curve with $\Omega$ of opposite signs,  but 
with absolute value matched smoothly through the gap (see Fig.~\ref{disc}),  
and use the conventional estimate for $D$. In other words 
we describe anomalous shear in region between the ring and the disc by the form of rotation curve. A problem 
here is that we may overestimate the width of the region in which 
the transition from the region co-rotating with
the disc to that co-rotating with the ring occurs. 
Contemporary observations fail to resolve this 
transition. In order to take this problem into account we consider
values of $D$ up to $D=200$, and demonstrate that the 
results are more or less generically stable.

\subsection{Simple estimates}

The basic concept of the standard galactic dynamo is that the azimuthal magnetic field $B_\phi$ is obtained 
from the radial
field $B_r$ by the action  of differential rotation $\Omega$,
while the radial field is restored from the azimuthal by the
$\alpha$-effect which arises from the mirror asymmetry of interstellar turbulence.
Both effects are in competition with  the destructive role of turbulent diffusion. This dynamo action can be represented in a very simple model by:

\begin{equation}
\frac{d B_{\varphi}}{d t}=r\frac{\partial \Omega}{\partial r} B_{r} - \eta(\pi^2/4) B_\varphi (1/h^2 +
 1/d^2),
\label{eqBphi}
\end{equation}
\begin{equation}
\frac{d B_{r}}{d t}= -\frac{\alpha}{h}
B_{\varphi} - \eta(\pi^2/4) B_{r}(1/h^2 + 1/d^2).
\label{eqBr}
\end{equation}
Here $\eta$ is the turbulent diffusivity, $h$ is the half-thickness of the disc and $d$ is the half-width of the ring.
For the crude estimates of interest, we have
replaced all derivatives in the dynamo equations by corresponding algebraic terms and introduced numerical factors $\pi^2/4$ to fit numbers known from numerical experiments (e.g. \cite{Phillips01}). We estimate magnetic diffusivity in the framework of
conventional mixing length theory as $\eta = lv/3$ where $l$ is the basic scale of turbulence and $v$ is the r.m.s.
turbulent velocity, and the numerical factor 3 comes from the dimensionality of space.  Another conventional estimate $\alpha = \Omega
l^2/h$ comes from the viewpoint that the mirror asymmetry of interstellar turbulence is caused by Coriolis force
action in a stratified medium.  Note that this conventional estimate implies that, for the case of a counter-rotating ring, $\alpha$ changes sign in the gap between the disc and the  counter-rotating ring.
This should be regarded as an order of magnitude estimate;
for our purposes the important factor is the dependence on $\Omega$.

\begin{figure}
\begin{center}
\includegraphics[width=9cm]{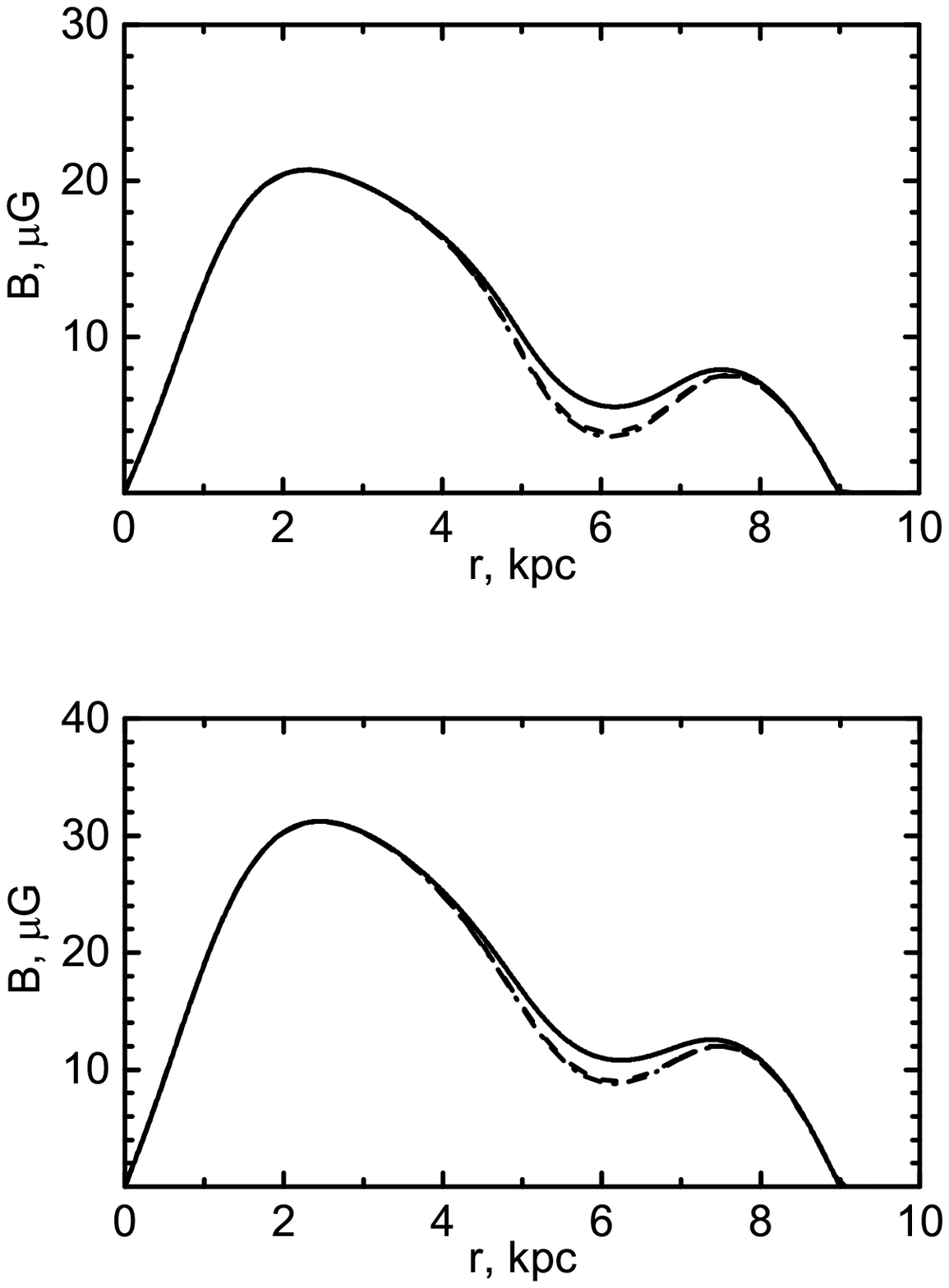}
\includegraphics[width=9cm]{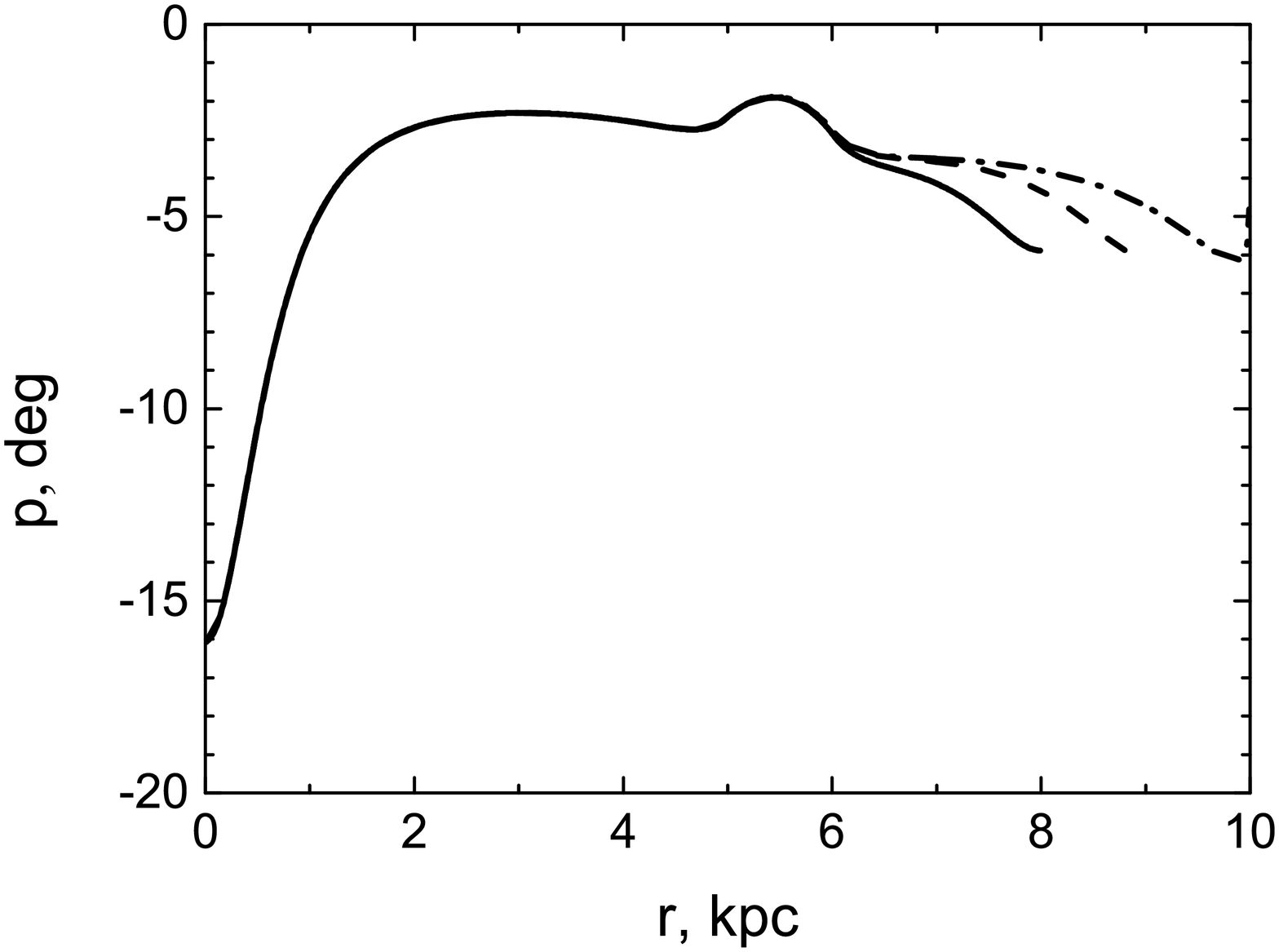}
\end{center}
\caption{Upper panel. Regular magnetic field generation  for various densities in the gap and the seed field in the inner disc. The 
upper panel shows the magnetic field at $t=10^{10} \mbox{ yr}$ for $D=20,$ the 
middle 
panel gives  the same for 
$D=50.$ The solid curve shows $\rho_{\rm gap}=10^{-1}\rho_{0}$, the dashed curve $\rho_{\rm gap}=10^{-2} \rho_{0},$
 and the dot-dashed curve  $\rho_{\rm gap}=10^{-3} \rho_{0}$. 
$R = 5$ kpc, $L=2$ kpc, $d=1$ kpc. 
Without a ring the outer maximum in $B$ is absent, and $B$ continues to decline with radius. Lower panel. The dependence of pitch angle $p=\tan^{-1}(B_r/B_\phi)$ on radius.}
\label{fig2b}
\end{figure}

Dynamo self-excitation means that the linear dynamo equations above have exponential solutions $\sim \exp \gamma t, \gamma>0$. (In particular, it means that all magnetic field components grow with the same rate.) After
some algebra we obtain from Eqs.~(\ref{eqBphi}), (\ref{eqBr}) a dispersion relation

\begin{equation}
\gamma = \frac{\eta}{h^{2}}\left(-(\pi^2/4)(1+ h^2/d^2) + \sqrt D \right),
\label{estimgam}
\end{equation}
where $D=-9 (h^2 \Omega  r\frac{\partial \Omega}{\partial r}/v^2)$ is a dimensionless quantity known as the dynamo number,
which measures the intensity of galactic dynamo action.
Dynamo self-excitation (positive $\gamma$) arises if $D$ exceeds
a critical level which for a thin galactic disc is $D_{\rm cr} \simeq \pi^4/16 \simeq 6$  (to be compared with $D_{\rm crit}=7.5$ 
found by \cite{Retal88} in a slightly different approach). For a flat rotation curve $|r
\frac{\partial \Omega}{\partial r}| \approx \Omega$ and for $h= 500 \mbox{ pc}$, $v=10 \mbox{ km s}^{-1}$ and 
rotation velocity of $200 \mbox{ km s}^{-1}$ we find that at $r=10 \mbox{ kpc}$, $ h^2 \Omega  r\frac{\partial 
\Omega}{\partial r}/v^2 =1$, i.e. $D=9$.

Diffusive losses in the gap between the ring and disc increase $D_{\rm cr}$. It reaches $D_{\rm cr} =10$
for $d = 1.5 h$, giving $d =600 \mbox{ pc}$ for $h=400$ pc. On the other hand, the above estimate are crude 
enough
that estimates $D = 10 \sim 50$ can be considered as realistic in a more 
complete model.
Increase of the ring thickness $h$ or
differential rotation $r \partial \Omega/\partial r$ promote dynamo self-excitation. Indeed, rotation curves of ring
galaxies appear more complicated than given by flat rotation curves (\cite{Silchenko14})
and the rings can be quite thick (\cite{uvrings}). We
conclude that self-excitation of a large-scale dynamo driven magnetic field in the rings looks possible. For $h =d$
dynamo excitation needs $D = 30$, which seems quite possible for the interstellar medium.

\begin{figure}
\begin{center}
\includegraphics[width=9cm]{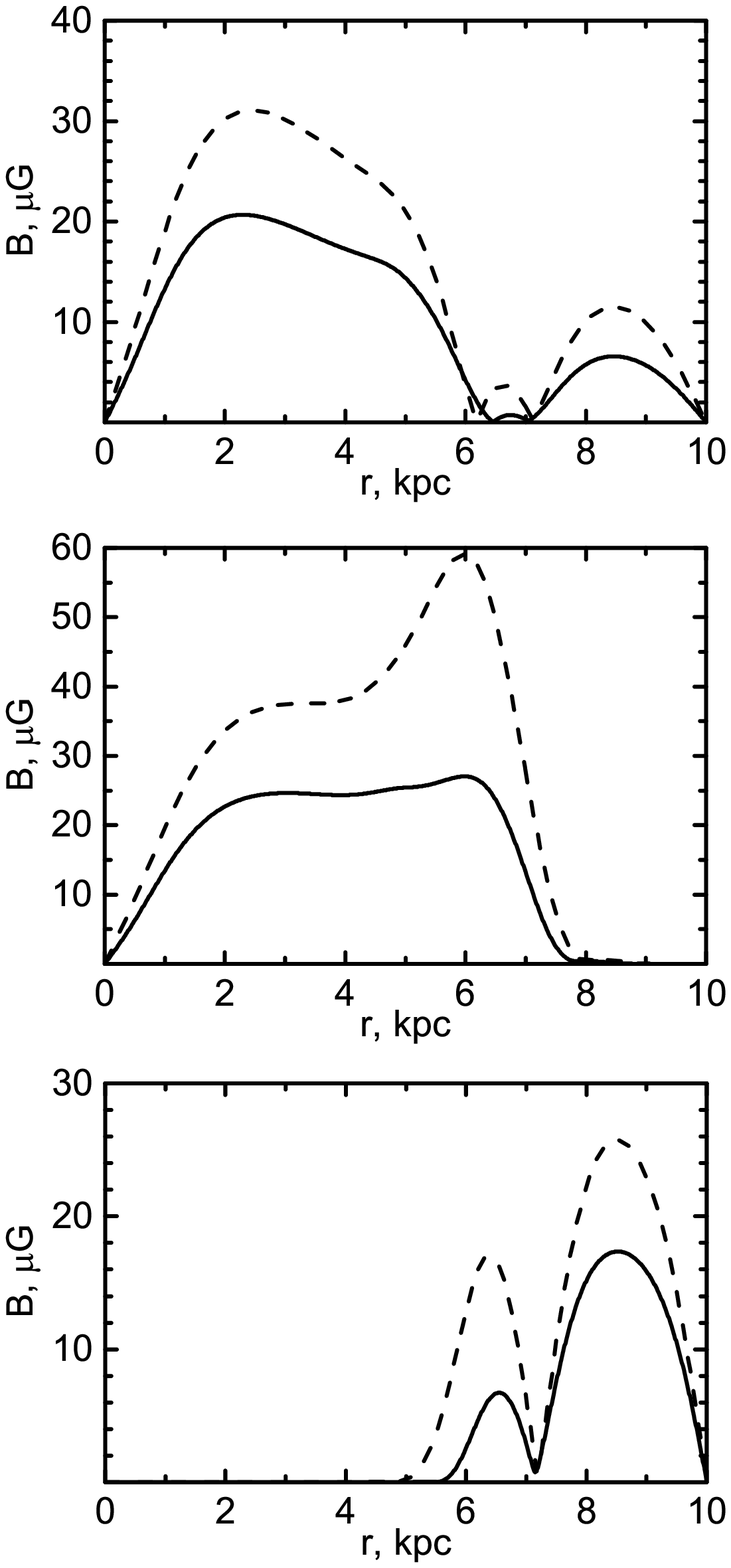}
\end{center}
\caption{Regular magnetic field generation for a counter-rotating ring
at a nominal time of $10^{10}$ yr . The solid curve shows the magnetic field 
for $|D|=20$, dashed curve for $|D|=50$. The upper panel shows $\alpha \propto \Omega$, middle panel  
$\alpha(r)=\alpha_{0}>0$, lower panel  $\alpha(r)=\alpha_{0}<0$. $R=5$ kpc, $L=2$ kpc.  
We note that in the lower panel the dynamo action in the disc is suppressed by the choice of sign of $\alpha_{0}$.
}
\label{fig2c}
\end{figure}

\subsection{Towards detailed dynamo models for ring galaxies}

A natural development of the above estimates is to use what is known as the no-$z$
approximation~(\cite{Subramanian93,Moss95}). This approximation describes the magnetic field
components in the disc plane. The $z$-component of the field perpendicular to the disc plane is restored by the
solenoidality condition assuming that it is much less than the plane components ($B_{r}$ and $B_{\varphi}$).
\begin{figure}
\begin{center}
\includegraphics[width=9cm]{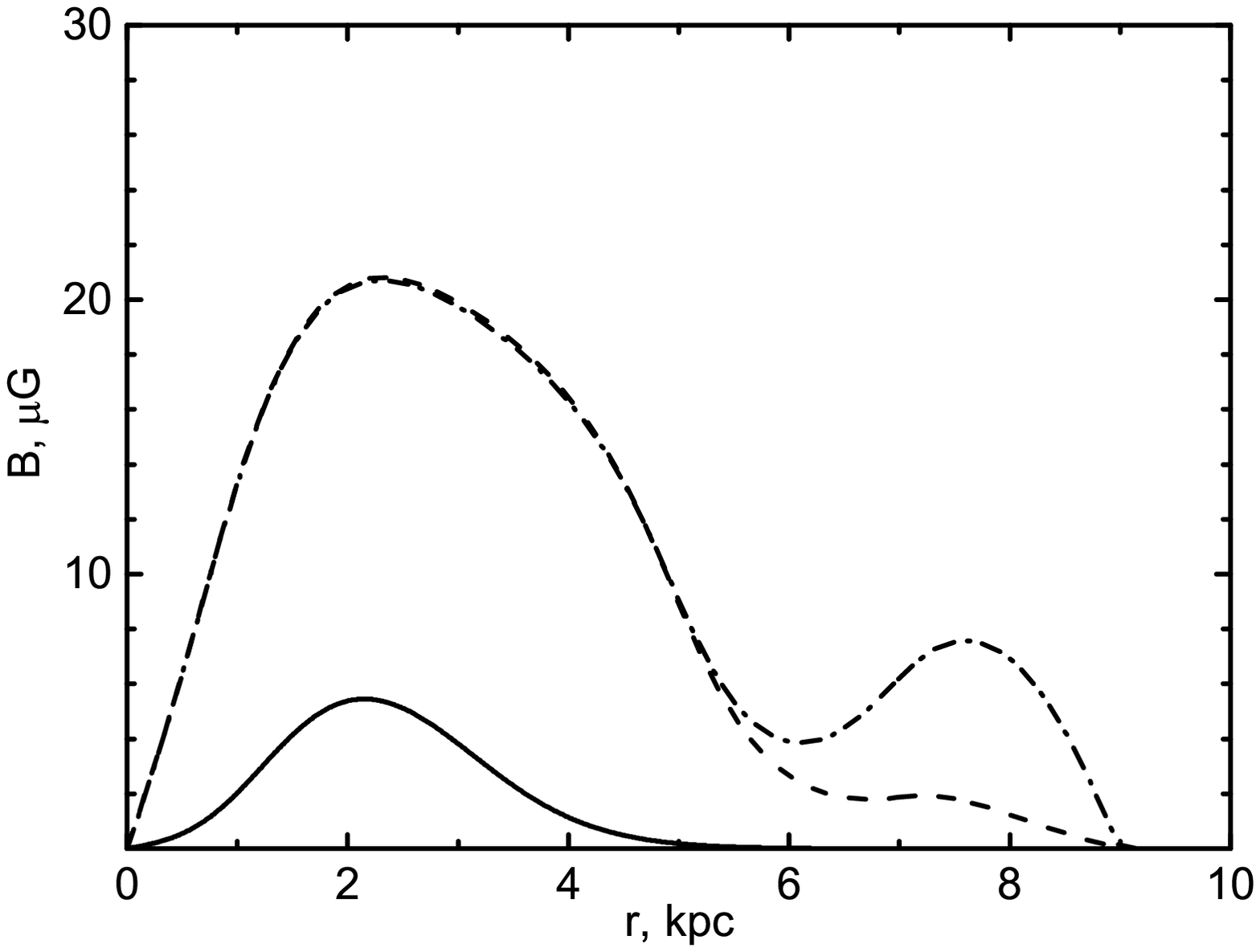}
\end{center}
\caption{Magnetic field evolution for the model with the seed field concentrated in the inner disc only ($R= 5$ kpc, $L= 2 $ kpc). The solid curve shows $t=2 \cdot 10^{9} \mbox{ yr},$ the dashed curve  $t=5 \cdot 10^{9} \mbox{ yr},$ and the
 dot-dashed curve is for  $t=10^{10} \mbox{ yr}.$}
\label{fig2a}
\end{figure}
In the framework of the no-$z$ model the dynamo equations read
$$\frac{\partial B_{r}}{\partial t}=-\frac{\alpha B_{\varphi}}{h}+$$
\begin{equation}\label{eq1}
+\eta \left(-(\pi^2/4)\frac{B_{r}}{h^{2}}+\frac{\partial}
{\partial r}\left(\frac{\partial}{r \partial r}(r B_{r})\right) +\frac{1}{r^{2}}\frac{\partial B_{r}}{\partial \varphi}\right),\end{equation}
$$\frac{\partial B_{\varphi}}{\partial t}=-r\frac{\partial \Omega}{\partial r} B_{r}-\Omega \frac{\partial B_{\varphi}}{\partial \varphi} $$
\begin{equation}\label{eq2}
+\eta
\left(-(\pi^2/4)\frac{B_{\varphi}}{h^{2}}+\frac{\partial}{\partial r}\left(\frac{\partial}{r \partial r}(r
B_{\varphi})\right) +\frac{1}{r^{2}}\frac{\partial B_{r}}{\partial \varphi}\right) \, .\end{equation}
The $\alpha$-coefficient includes saturation of the magnetic field growth when it reaches the equipartition value
$B^{*}=v\sqrt{4\pi\rho}$ where $\rho$ is the interstellar medium density. We use an algebraic quenching for dynamo action  $\alpha \propto (1+\frac{B^{2}}
{B^{*2}})^{-1}$ and avoid for the time being more sophisticated parameterisations
~(e.g., \cite{Shukurov06,Sur07,Mikhailov13}). For $\Omega$ we use a Brandt (1960)  rotation curve:

\begin{equation}
\Omega=\Omega_{\rm Br}(r)=\frac{\Omega_{0}}{\sqrt{1+(r/r_{\omega})^{2}}},
\label{Brandt}
\end{equation}
with parameters chosen to give an asymptotic rotational
velocity of about $200$ km s$^{-1}$ when $R_{\omega}=2\mbox{ kpc}.$ 
Now  $\alpha\propto\Omega$.

We assume that $B^{*}=5 \mbox{ } \mu \mbox{G}$ in the main parts of the
galaxy. 
For $\rho_{\rm gap}$ we
assume that it is an order of magnitude less than the density in the inner parts
and $B^*$ is then about 0.5 $\mu G$ there (Fig.~2, lower panel). 
Usually we use zero boundary conditions at the central and outer boundaries of the galactic disc. 
Additionally, we define $L$ to be the radial extent of the gap between 
inner disc and outer ring, and $R$ the radius of the inner disc. 
As a conservative assumption we take the rotation curve to be given 
by Eq.~(\ref{Brandt}) everywhere, including in the gap.
The outer boundary of the disc in which we model dynamo action is fixed
at a radius of 10 kpc in all cases.

The magnetic field configuration in a ring galaxy depends on the lifetime of the ring and its origin, in 
particular the nature  of the seed magnetic field configuration. The main aim of this paper is to investigate whether a galactic 
dynamo can generate a magnetic field in the ring. 
In order to clarify this, we are interested initially in 
long-lived rings and weak (much smaller than equipartition strength) seed magnetic fields. 
This is why we choose as a basic example to present the field resulting from a weak seed field after $t=10^{10}$ yr.
We bear in mind that other possibilities should be  considered later, 
after confirming that dynamo generated magnetic 
fields can be important for ring galaxies. 
Generally the saturated state is attained in a shorter time, but at the 
moment we wish mainly to demonstrate the eventual outcome of dynamo action. 
In the framework of this research 
we first perform some preliminary 
experiments.

The modelling is quite conservative, in that
we use the Brandt rotation curve and ignore the fact that
differential rotation in ring galaxies is usually more pronounced,
and always consider an inner disc of maximum radius $R= 5$ kpc
independent of the width of the gap $L$.
Correspondingly,
an increase of $L$ moves the ring to a less dynamo-active radius.

\section{Results: no-$z$ model}
\label{res}

We have run our model for $D=9,\mbox{ }20, \mbox{ }50$ and
some other values that seem to be realistic for spiral galaxies
and their outer rings. For $D=9$ the initial magnetic field in the
ring decays. Fig.~2 shows $B=\sqrt{B_{r}^{2}+B_{\varphi}^{2}}$ for
$t=10^{10} \mbox{ yr}$ for $D=20$ (top panel) and $D=50$ (middle
panel) and the seed field present in the outer ring only
($B_{\varphi}(t=0)=B_{0}\sin \left(\pi \frac{r-r_{\rm
min}}{2d}\right)$, $B_r =0$ and $r_{\rm min}$ and $r_{\rm max}$
are the inner and the outer radii). We find that the intensity of
dynamo action ($D$) can be (under realistic assumptions) sufficient for magnetic field
self-excitation and that the field can reach a field strength  that is close to
the equipartition value in a few Gyr. 
Varying the  half-widths of the ring from 1 to 2 kpc
(the magnetic field decays for smaller $d$) we find that the
magnetic configuration is 
quite robust. However wider rings
and larger $D$ give more widely distributed and stronger magnetic field.
In order to isolate the effects of dynamo action in the ring we take here  
a seed magnetic field for the dynamo that is present in the ring only.
We also can conclude that an initial magnetic field concentrated in
the ring can pass into the central parts.

\begin{figure}
\begin{center}
\includegraphics[width=9cm]{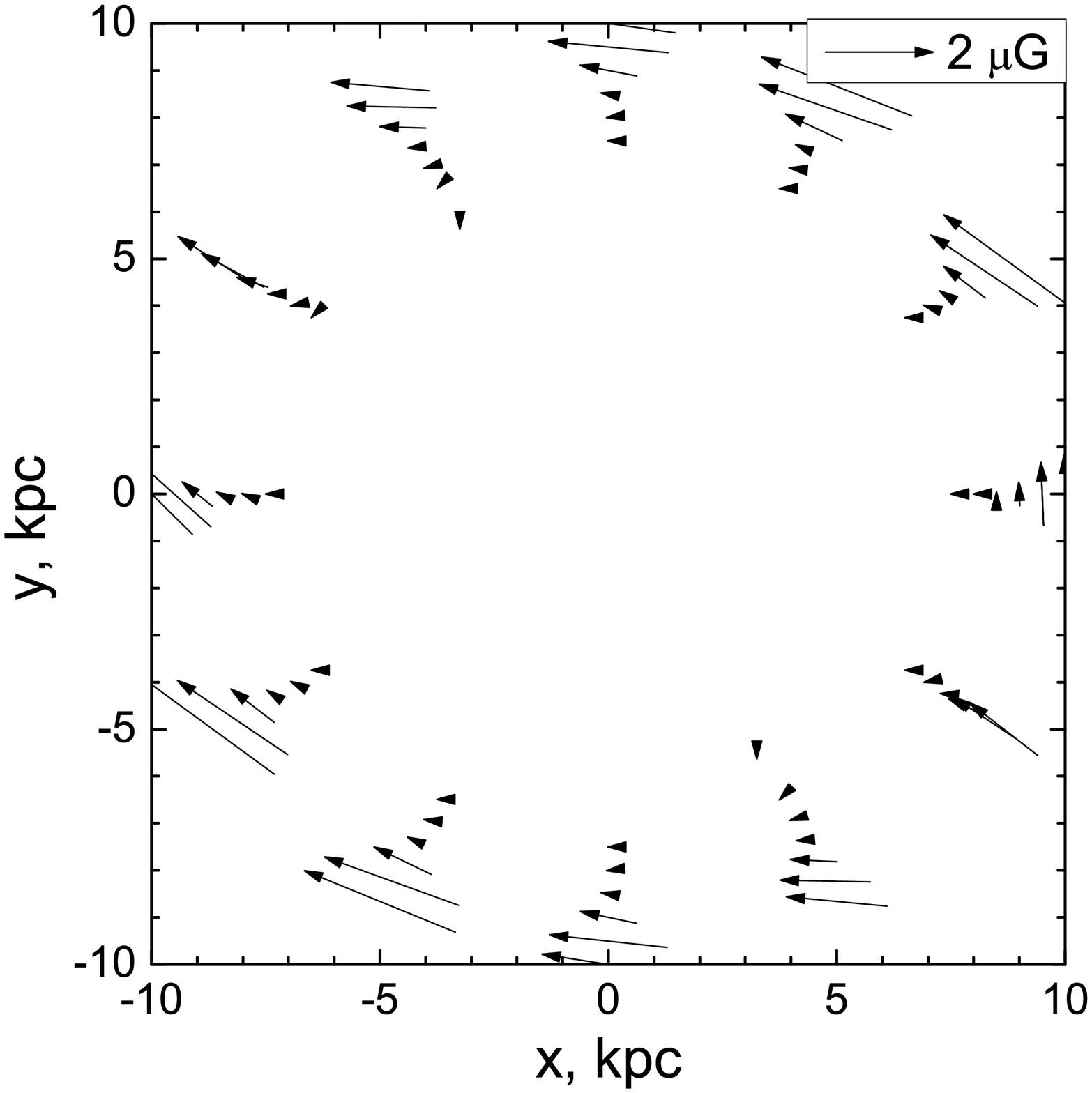}
\end{center}
\caption{Regular magnetic field generated in an outer thin ring} in the nonaxisymmetric case  (see discussion  in the text).
\label{fig2d}
\end{figure}

The figure obtained  for the  large-scale magnetic field strength in the disc is in some cases as large as $30 \, \mu G$ which appears
rather large in the context of magnetic fields in spiral galaxies. This estimate 
arises because we use $B_{\rm 
eq} = 5 \, \mu G$ and $D=50$. Then, using the estimate $B = B_{\rm eq} \sqrt{D-D_0}$
obtained by Shukurov (2007),  where $D_0 =7$ is the generation 
threshold for the disc, we obtain the above field strength.  

\begin{figure}
\begin{center}
\includegraphics[width=0.45\textwidth]{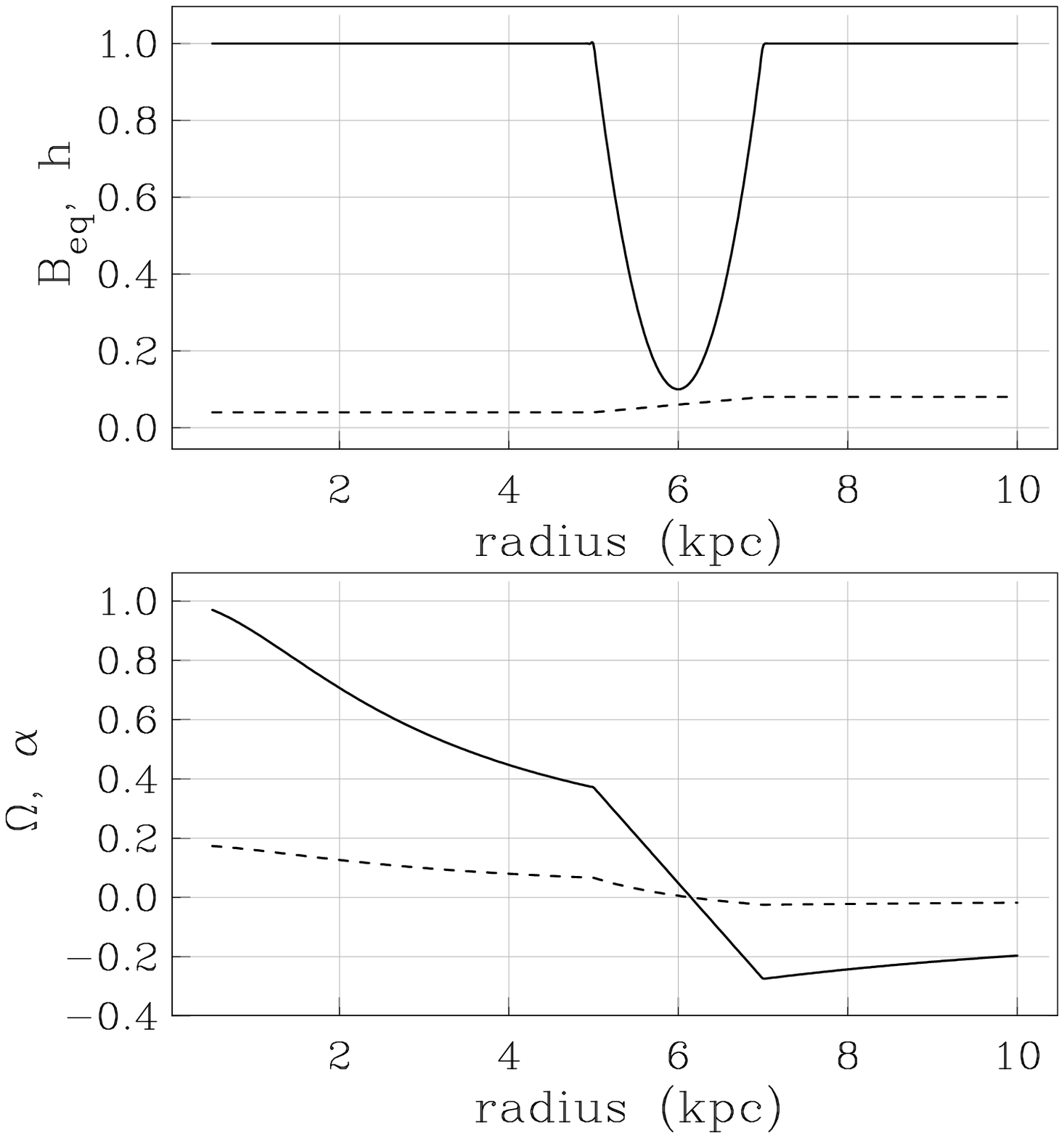}
\end{center}
\caption{ Radial dependence of the key disc properties for the
counter-rotating model. Upper panel, $B_{\rm eq}$ (solid, in units of $ B^*$). 
Lower panel $\Omega$ (solid) and $\alpha$ (broken. }
\label{disc}
\end{figure}

Then we ran the model with the seed in the inner disc only (Figs.~3 and 4). We see that the outer ring becomes magnetized
after sufficient time, but it appears that magnetic field transport from the inner disc is at least as
important as in situ dynamo action in the outer ring.

Fig.~5 shows that  in the framework of our model
the dynamo generated magnetic field is not very sensitive to the 
gas density in the gap between the ring and the disc. 
Without a ring, the outer maxima in $B_\phi$ are absent, and $B_\phi$ continues with radius.

The no-$z$ model gives a counter-intuitive result (Fig.~6) for the case of a counter-rotating ring
(such as observed in NGC~4513, although we do not attempt to model this galaxy explicitly). Then $\Omega=f(r)\Omega_{\rm Br}(r)$
where $f(r)=1$ in the inner parts, $f(r)=-1$ in the outer parts and in the gap it changes gradually. Qualitatively, $\Omega(r)$ is as shown in
Fig.~\ref{disc} , with allowance for differing positions of the gap.
Counter-rotation of the ring in respect to the inner disc means very large radial gradients in $\Omega$. A
natural expectation is to get effective dynamo action around the gap between the ring and the disc. In practice,
we face a problem how to parametrize in the framework of the model the quantity $\alpha$ for the counter rotating
ring. A straightforward parametrization $\alpha=\Omega l^{2}/h$ assumes that $\alpha$ vanishes in the
gap, suppressing dynamo action. (Here $l$ is the scale of interstellar turbulence: typically $l=100$ pc.) $\alpha$ 
is a quantity which is hard to predict, and we also tried the option that $\alpha$  does not change sign in
the gap (Fig.~6c). In the absence of counter-rotation, we tested the 
case of $\alpha$ positive in the disc as well as in the 
ring as well as the case of
$\alpha$ negative both in the disc and the ring
and found  again that the
dynamo action in the gap region  is quite moderate.

\begin{figure}
\begin{center}
a)\includegraphics[width=0.37\textwidth]{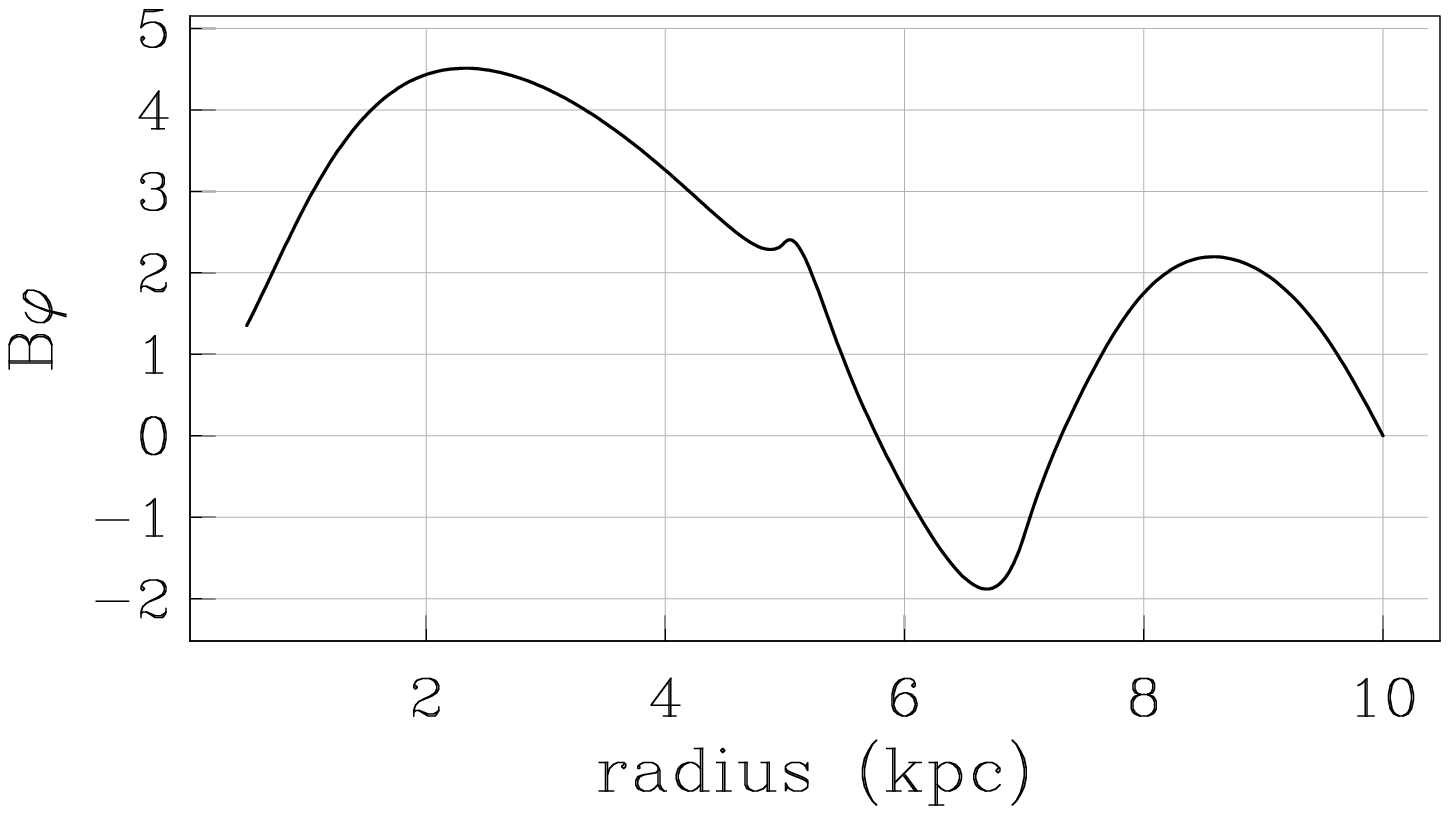}\\
b)\includegraphics[width=0.45\textwidth]{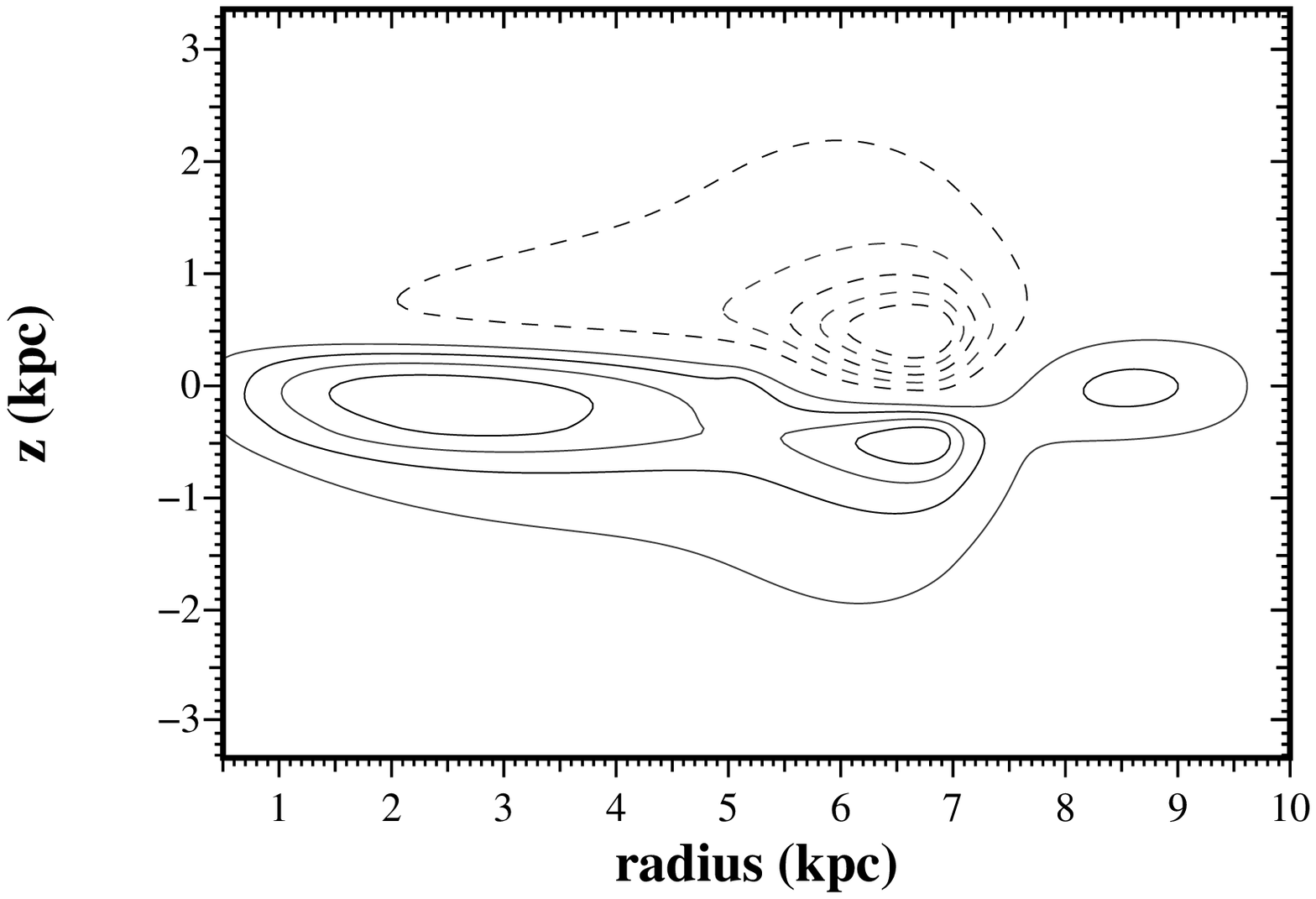}\\
\end{center}
\caption{a) Dependence of $B_\phi$ on radius in the disc plane,
b) $B_\phi$ contours for the model with $B_{\rm eq1}=0.1 B^*$ ; Here
$r_i=0.5R_{\rm gal}, r_o=0.7R_{\rm gal}$, with counter-rotation.} \label{2D2}
\label{2D2fig}
\end{figure}

We see from the above results that the magnetic field can pass through the gap between ring and disc. 
If the seed is
non-zero in the disc only, it penetrates into the ring and vice versa.
A theoretical prediction for the field propagation (based on an estimate given in a much more general context in
\cite{Kolmogorov37}) is given by  (\cite{Moss98,Mikhailov15}):

$$V=\sqrt{2 \gamma \eta},$$
where  $\gamma$ is the growth rate of magnetic field. Taking
$\eta=0.33 \mbox{ kpc km s}^{-1},$ $\gamma=1.5
\mbox{ Gyr}^{-1},$ we obtain
$V=1 \mbox{ kpc Gyr}^{-1}$, quite close to the
numerical estimate that follows from Fig.~4.
We have assumed that the diffusivity, given by properties of the
turbulence is uniform through the inner disc and ring. 
Of course, this may not be justified.

We recognize that these times significantly exceed the probable lifetimes 
of the rings. Among other conservative assumptions in the model, we note
that the timescales depend inversely on the value of the diffusivity, which is
subject to substantial uncertainty. Also the choice of seed field and 
the history of the galaxy before the formation of the ring can be important -- see the
comments in Sect.~\ref{2D}.

\begin{figure}
\begin{center}
a)\includegraphics[width=0.45\textwidth]{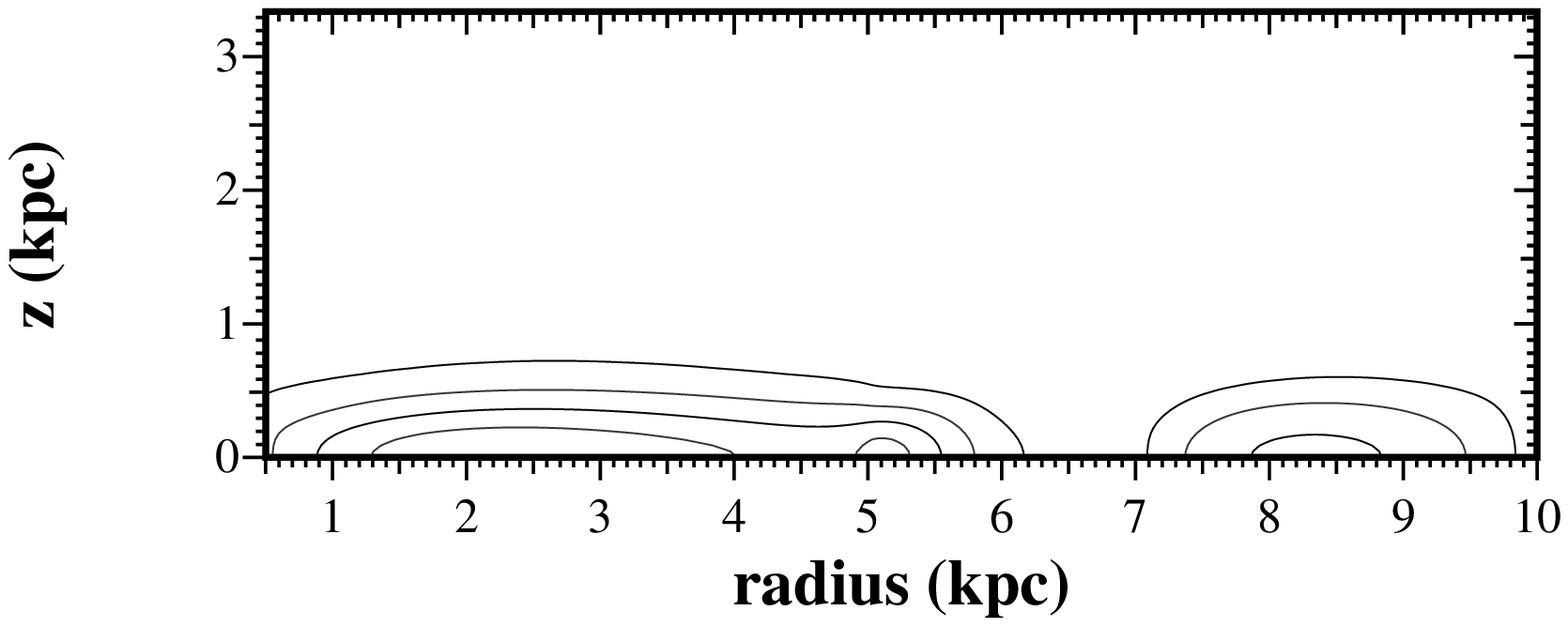}\\
b)\includegraphics[width=0.37\textwidth]{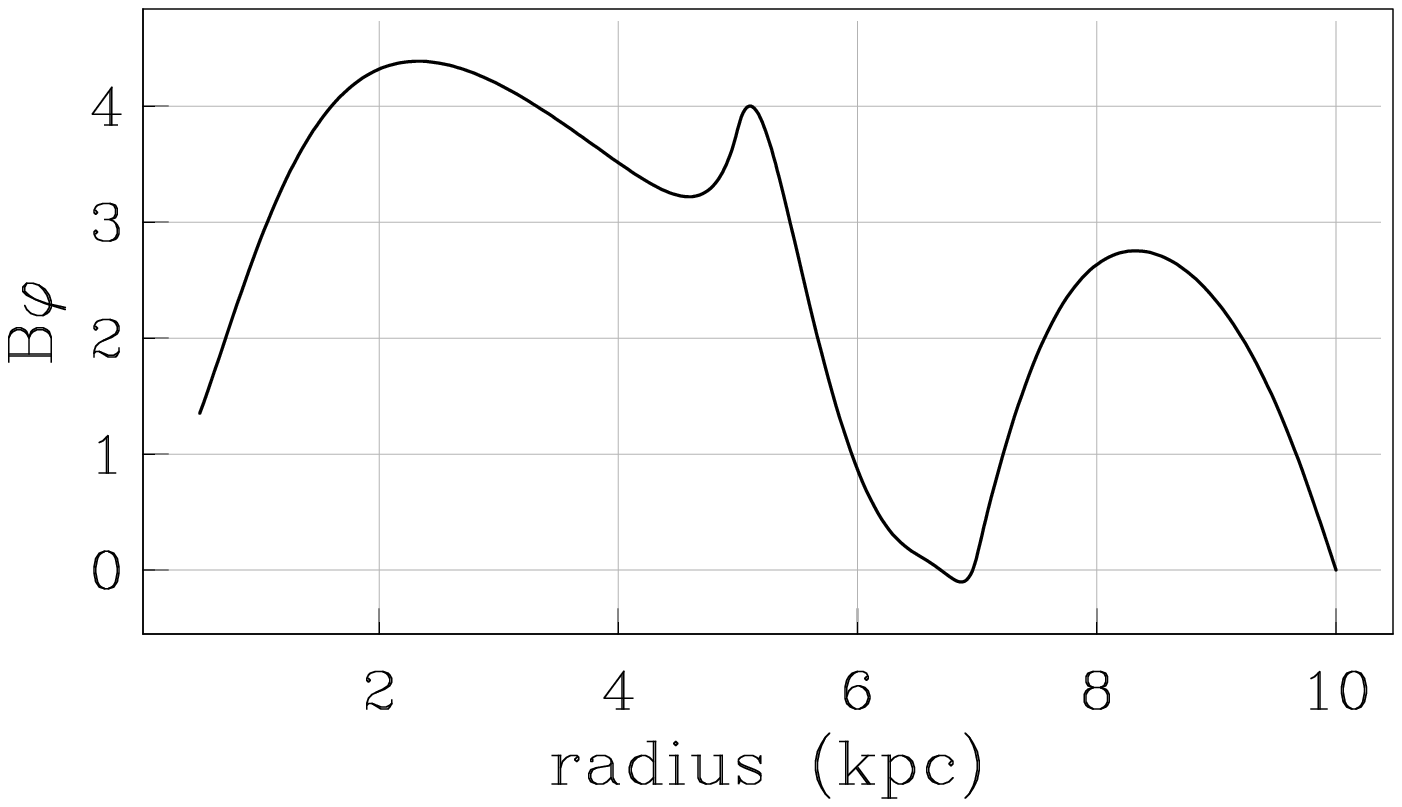}\\
c)\includegraphics[width=0.37\textwidth]{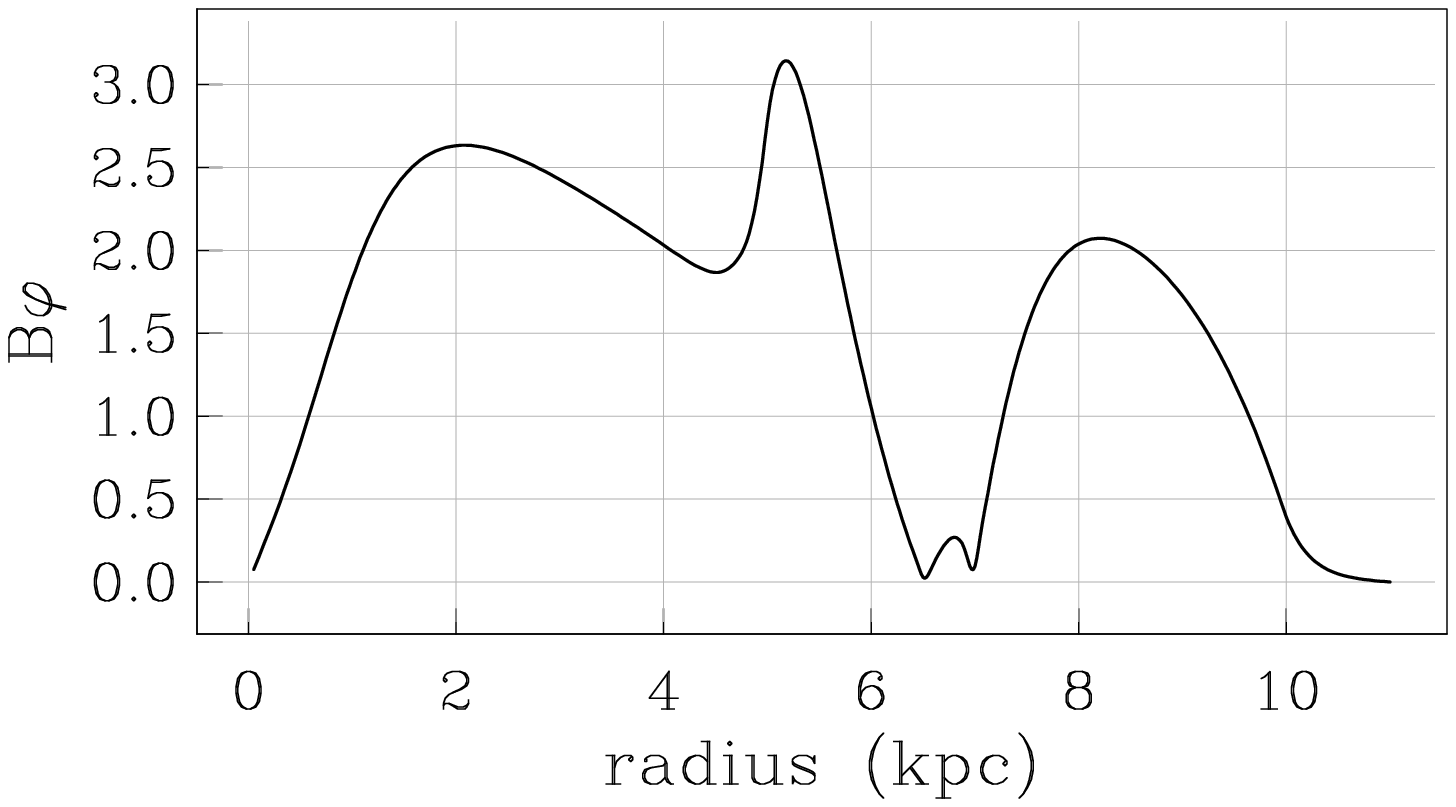}
\end{center}
\caption{a) Dependence of $B_\phi$ on radius in the
disc plane; b) $B_\phi$ contours, for model with $B_{\rm eq1}=0.1B^*$
and enforced quadrupolar parity; c) Dependence of $|B_\phi|$ on radius
for the corresponding no-$z$ model. $r_i=0.5 R_{\rm gal}, r_o=0.7 R_{\rm gal}$, with
counter-rotation. $B_\phi$ is measured in units of the equipartition field $B^*$.} \label{2D10}
\end{figure}

The magnetic field in an isolated galactic disc is usually (almost) axisymmetric. 
A ring however is a relatively thin,
azimuthally extended and
relatively isolated, body and it is  a priori possible that generation 
of a bisymmetric magnetic field could occur there.
We can note here that dominant bisymmetric magnetic configurations
are now considered to be rare in or completely absent from normal spiral galaxies
(see e.g. Beck 2016). 
Nevertheless, it seems reasonable to examine briefly this option in thin rings,  given that
dynamo theory suggests that thin rings
are believed to be more favourable for generation of nonaxisymmetric fields.
However, we found such a solution only for the case of a free outer boundary of the disc ($\partial B/\partial r=0$ at
$r=10$ kpc), $\Omega_{0}=67 \mbox{ km s}^{-1} \mbox{kpc}^{-1}$, $\alpha=2.68 \mbox{\, km s}^{-1}$ (corresponding to $R_\alpha\approx 4$, $R_\omega \approx 50$)
the radius of the inner disc is $5 \mbox{ kpc},$ the inner radius of the outer ring is $9 \mbox{ kpc}$, and its outer 
radius is $10 \mbox{ kpc}$. 
The initial field is zero if $r<9\mbox{ kpc}$ and bisymmetric in the ring.
With the standard dynamo parameters of Sect.~4, the bisymmetric field generated
only survives for about 
$t=0.5 \mbox{ Gyr}$  and then disappears, to be replaced by an axisymmetric configuration. However with 
slightly different dynamo parameters a stable
bisymmetric field can be found (Fig.~7).
We can conclude that, although possibly of some interest in a dynamo theory 
context, nonaxisymmetric fields are not likely to be relevant to real ring galaxies.

\begin{figure}
\begin{center}
\includegraphics[width=0.45\textwidth]{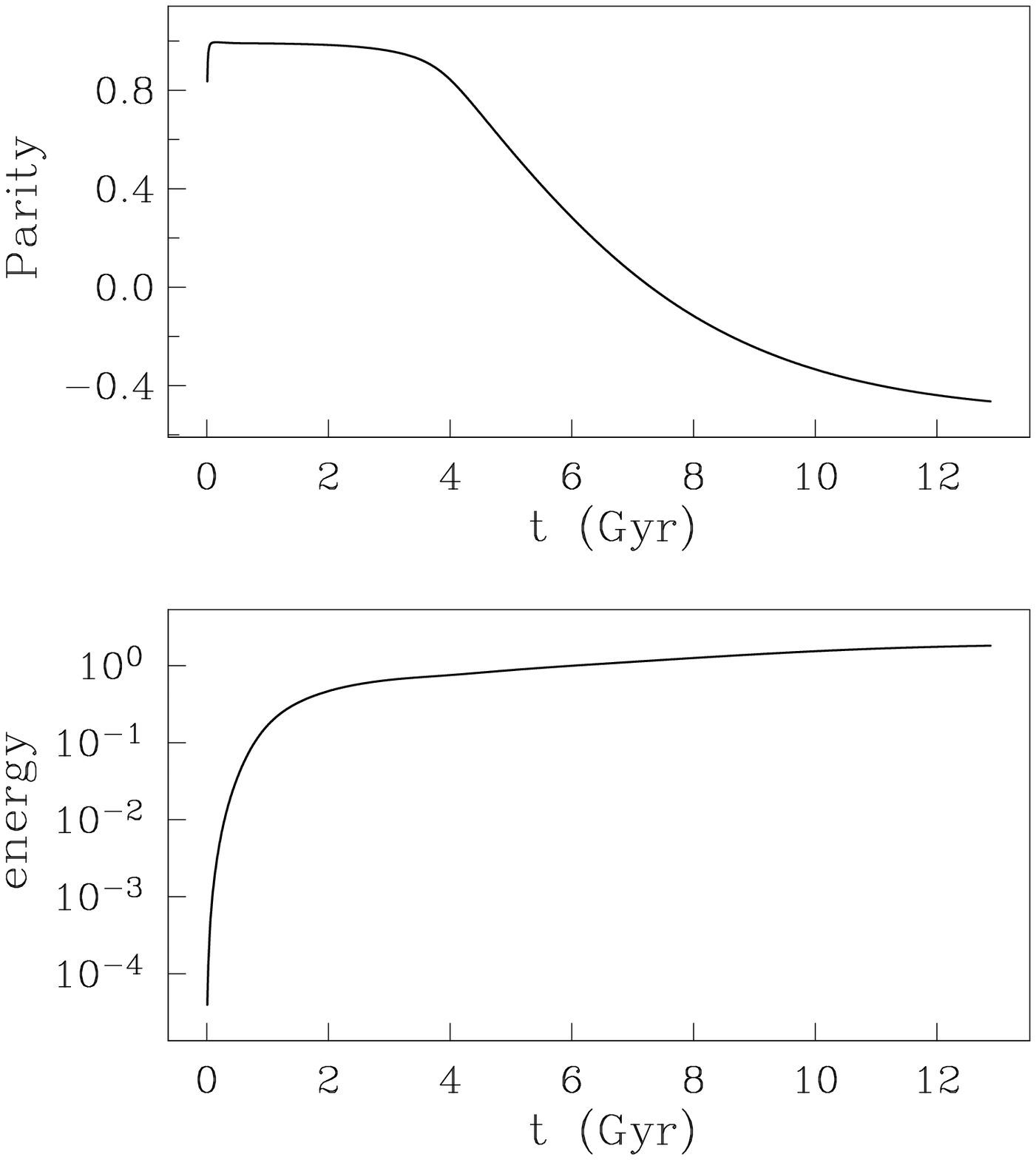}
\end{center}
\caption{Dependence of global parity and energy on time for
a calculation beginning from a weak seed field of mixed parity.
} \label{evol_seed}
\end{figure}

\section{Counter-rotating discs -- beyond the no-z model}
\label{2D}

\subsection{Basic model}
\label{basic2D}
The rather surprising results obtained above for counter-rotating discs,
using the severely truncated no-$z$ model, suggests the need for verification
with an alternative model. Of course, such verification would be 
helpful for case of a co-rotating disc as well, but in this case previous
comparisons show that the no-$z$ model provides a satisfactory
approximation to a conventional axisymmetric model with explicit $z$-dependence. Given that the no-$z$ models of Sect.~\ref{res}
suggest that fields are axisymmetric, we revisited the problem using an
axisymmetric embedded disc model, with cylindrical $r, z$ coordinates.
The code used is essentially that of Moss et al. (1998).
(This is an $\alpha^2\Omega$ code, whereas in the previous sections the
$\alpha\Omega$ approximation is used. However, in the regime of interest,
there is little difference between these approaches.)
In this Section we use a slightly different notation: $R_{\rm gal}$ is now the overall galactic radius, and we define the gap location and size by parameters $r_i$, $r_o$.

The reference rotation curve is again a Brandt curve (Eq.~(\ref{Brandt}))
with $r_\omega=0.2R_{\rm gal}$. In this $\alpha^2\Omega$ code,
the dynamo numbers are given
by $R_\omega=\Omega_0h^2/\eta_0, R_\alpha=\alpha_0 h/\eta_0$, where $\Omega_0$ is 
the quantity defining the rotation curve (eq.~(\ref{Brandt}), 
$\alpha_0$ is the corresponding value of $\alpha$, and $\eta_0$ is the value of the diffusivity
in the disc. 
The diffusivity is independent of radius, and increases asymptotically to
$25\eta_D$ in the halo.
$\alpha$ varies sinusoidally in the disc ($\propto \sin\frac{z\pi}{h(r)}$),
and is zero in $|z|>h(r)$.
We continue to use a canonical value of $\eta_D=10^{26}$ cm$^2$s$^{-1}$ in the later
discussion, but always bearing in mind the possibility of some variation, 
as this estimate is rather uncertain. Our probable underestimate of the angular velocity gradients, and 
hence
effective dynamo number, might be considered to be compensated by the use of large values of $D$.
The "gap" is between $r=r_i$ and $r=r_o$,
and the equipartition field strength is uniform outside of the gap, i.e.
in $r\le r_i$ and $r\ge r_o$. Field strengths are measured in units of
this equipartition field (again defined by the kinetic energy
of the assumed turbulent gas motions). In the gap the equipartition field
drops quadratically to
the value $B_{\rm eq1}$ at $r=(r_i+r_o)/2$. In the initial investigation, 
the dynamo parameters were
slightly different to those used in Sect.~\ref{res}: $R_\alpha=4$,
$R_\omega=50$. For the majority of models,
in $r\le r_i$, the disc thickness $h=0.04R_{\rm gal}$,
in $r\ge r_o$, $h=0.08R_{\rm gal}$, with a smooth interpolation in the gap.
Analogously, in $r\le r_i$, $\Omega=\Omega_B(r)$,
in $r\le r_o$, $\Omega=-\Omega_B(r)$, with a smooth transition.
Correspondingly, with $\alpha\propto \Omega$, $\alpha$ changes sign smoothly.
In the standard case, $r_i=0.5 R_{\rm gal}, \,  r_o=0.7 R_{\rm gal} \ $. 
Fig.~\ref{disc} shows the radial dependence of the key disc quantities.

We first computed a model with $B_{\rm eq1}=0.1 B^*$ (Sect.~3.2). Fig.~\ref{2D2}a
shows the dependence of $B_\phi$ on radius along the mid-plane $z=0$,
and Fig.~\ref{2D2}b  the contours of $B_\phi$.
The notable feature of these plots is the departure from strict quadrupolar
parity -- in this case parity $P\approx -0.5$ in the steady state.
This feature could not be revealed by the no-$z$ model, which implicitly
assumes $P=+1$.
We have used the standard definition of dynamo field parity,
\begin{equation}
P=\frac{E_Q-E_D}{E_Q+E_D},
\end{equation}
where $E_Q, E_D$ are the energies of the even and odd parts respectively
of the field. Thus a strictly quadrupolar/dipolar field has $P+1/-1$.

In order to make a better comparison with the no-$z$ model, we recomputed
this model in the region $z\ge 0$, with quadrupolar parity enforced
by the boundary conditions on the plane $z=0$.
The contours of toroidal field and the radial variation of $B_\phi$
on $z=0$ are shown  in the upper two panels of Fig.~\ref{2D10}.
The lower panel shows the variation of $|B|$ with radius in
the corresponding no-$z$ model.
Comparison of the lower two panels of this Figure shows a remarkable similarity
between the no-$z$ and $r,z$ models, and gives confidence in the modelling,
even for these much larger dynamo numbers.
(See also Phillips 2001).

Varying $B_{\rm eq1}$ between 1 and 0.01 produces very little variation in the eventual steady configuration,
with final parity around $P\approx -0.3$.
Moving the gap to $(r_i, r_o)=(0.6, 0.9)$ gives a final parity, $P\approx +0.4$,
and with (0.35, 0.7) gives $P\approx -0.28$.
A model with uniform disc thickness, $h/R_{\rm gal}=0.04$, no gap/counter-rotation,
$\alpha\propto \Omega$ and $R_\alpha=-1.0, R_\omega=20$ gives a strictly dipolar field,
$P=-1$. This result persists for considerably larger dynamo numbers, e.g. $R_\alpha=-4.0, R_\omega=50$. This means that the dynamo action in the disc of the model is strong enough to excite a dipolar
magnetic field  (in which the azimuthal field changes sign at the central plane (antisymmetric)) while for a weaker dynamo action magnetic 
fields in galactic discs have a quadrupolar
configuration  (with azimuthal field that is symmetric with respect to the central plane). The counter-rotating ring and 
associated magnetic field losses resulting from magnetic field transport into
the gap makes the parity of the configuration mixed, and the field configuration in the inner parts of the disc is 
closer to 
quadrupolar symmetry.

\begin{figure}
\begin{center}
(a)\includegraphics[width=0.45\textwidth]{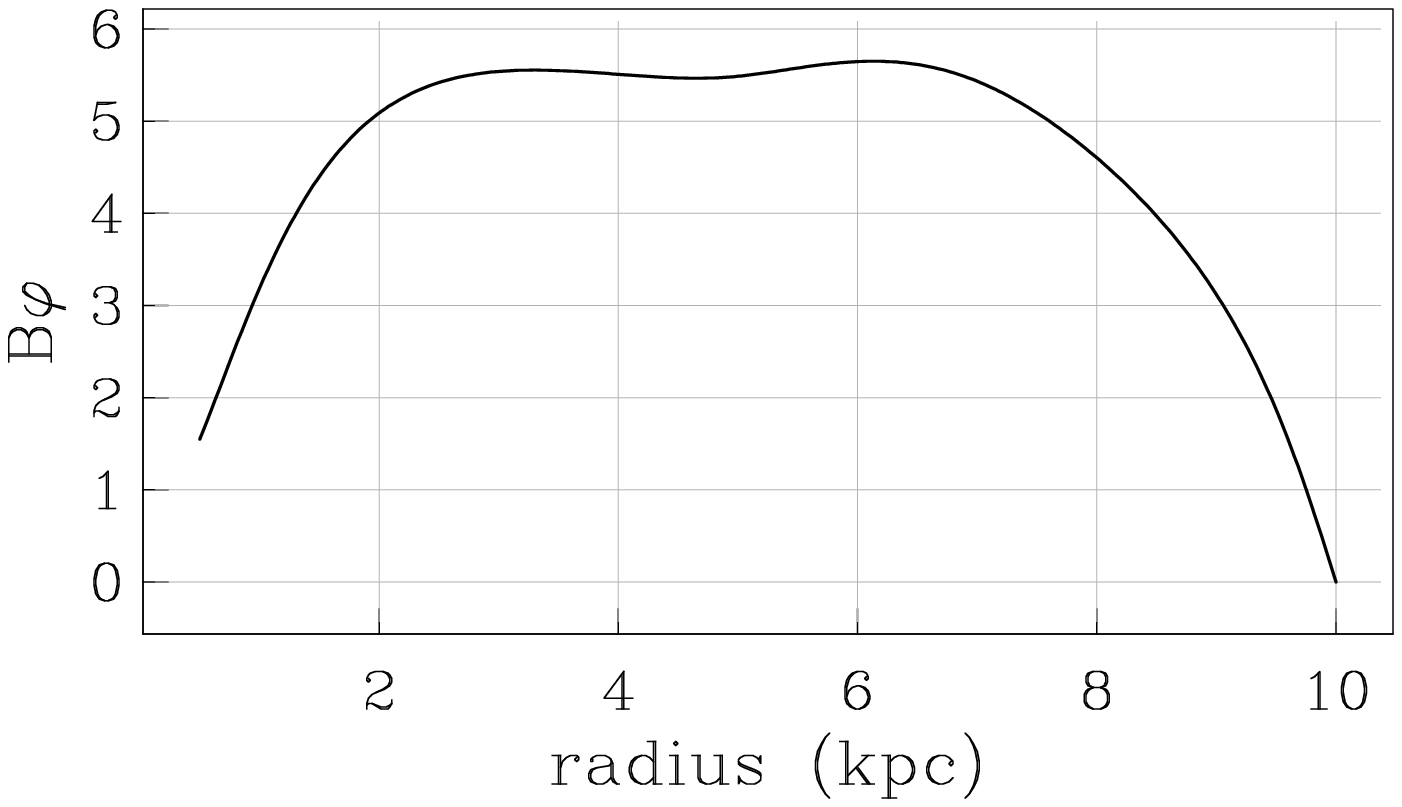}\\
(b)\includegraphics[width=0.45\textwidth]{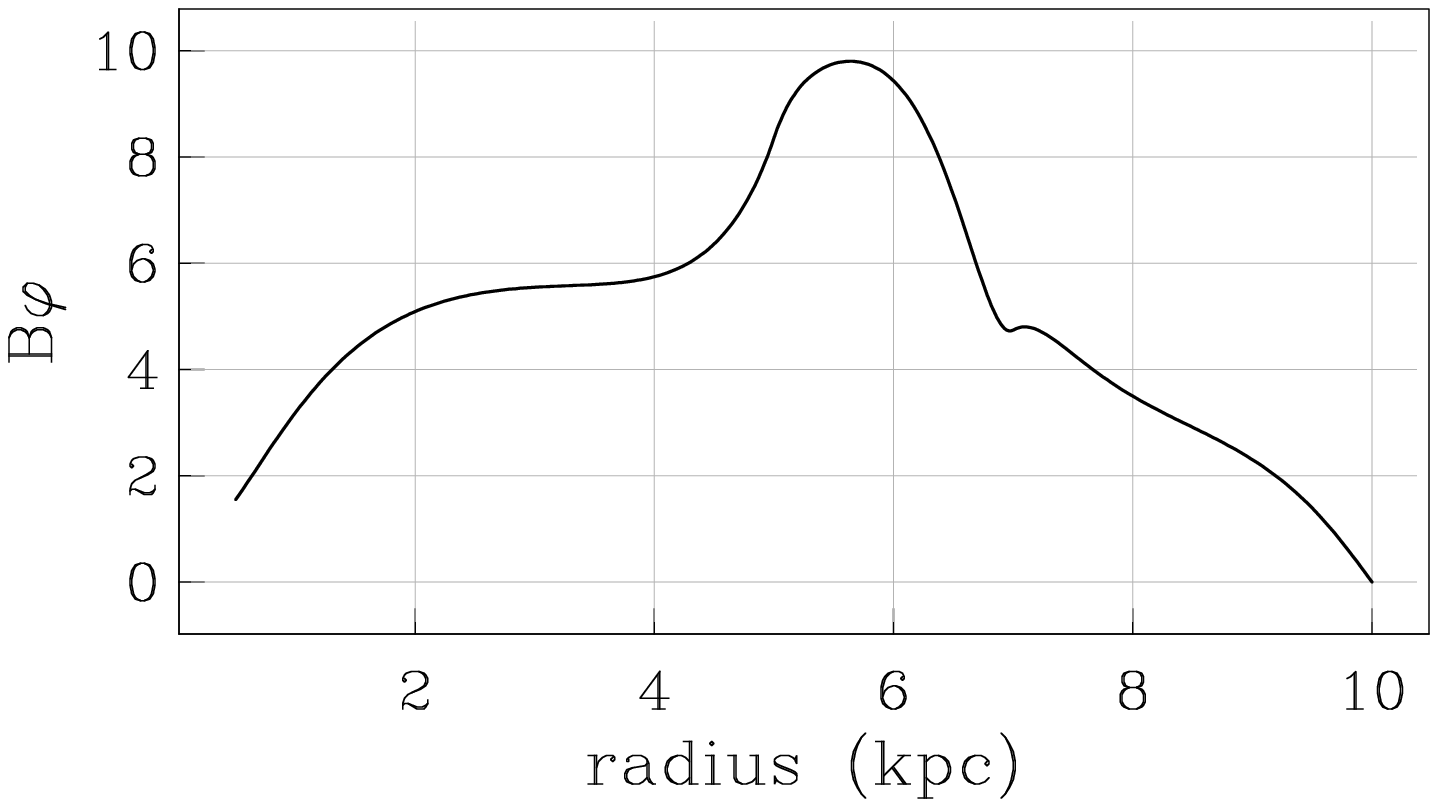}\\
(c)\includegraphics[width=0.45\textwidth]{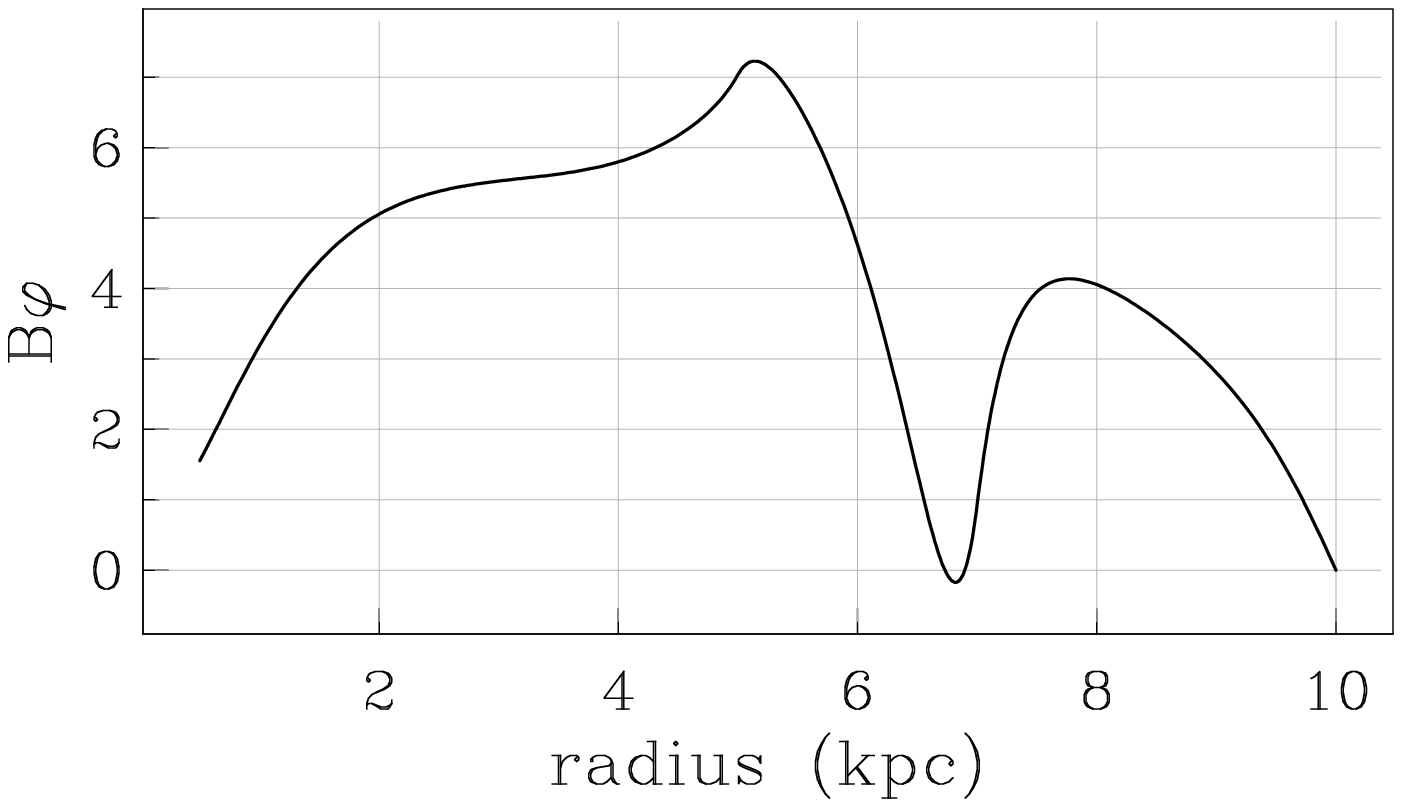}\\
(d)\includegraphics[width=0.45\textwidth]{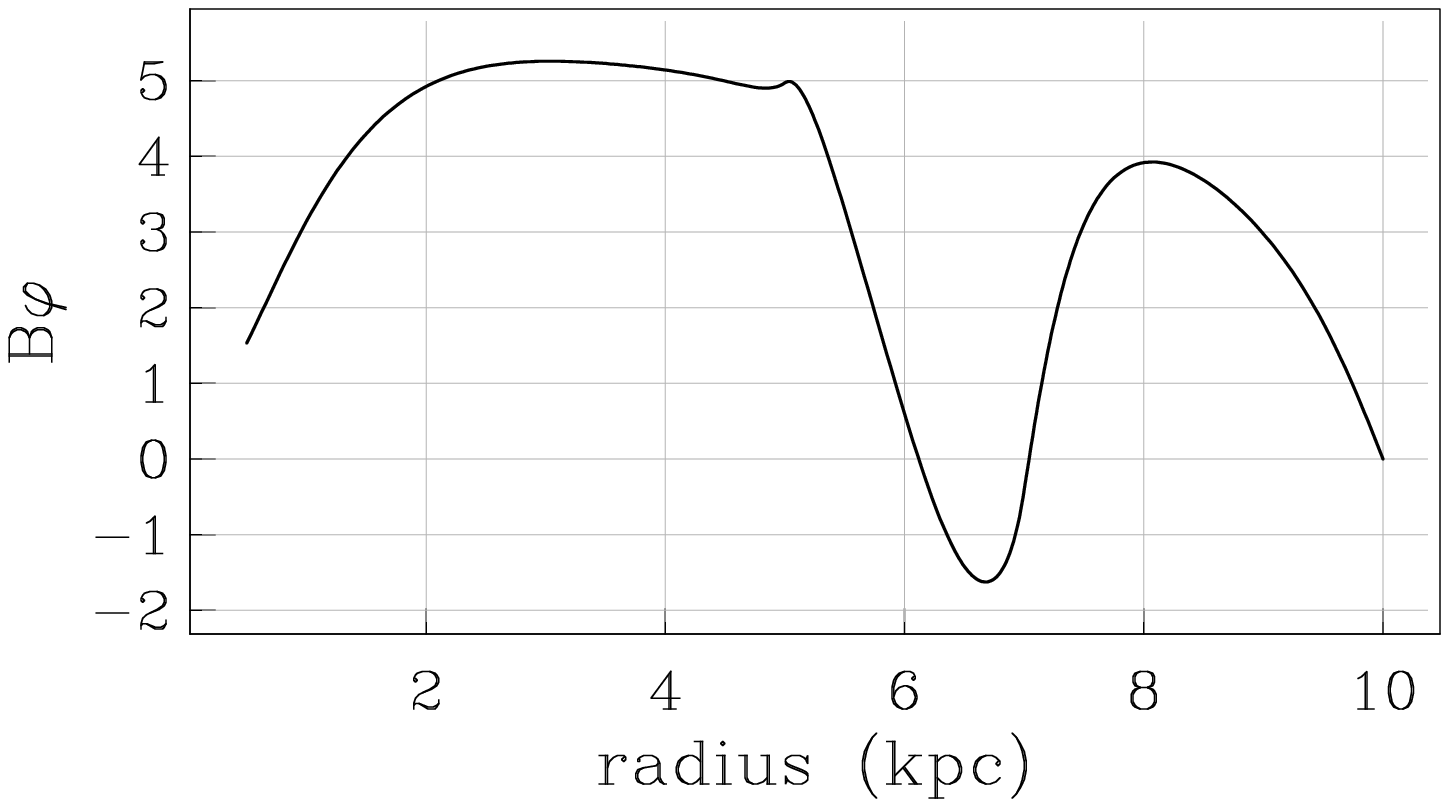}\\
(e)\includegraphics[width=0.45\textwidth]{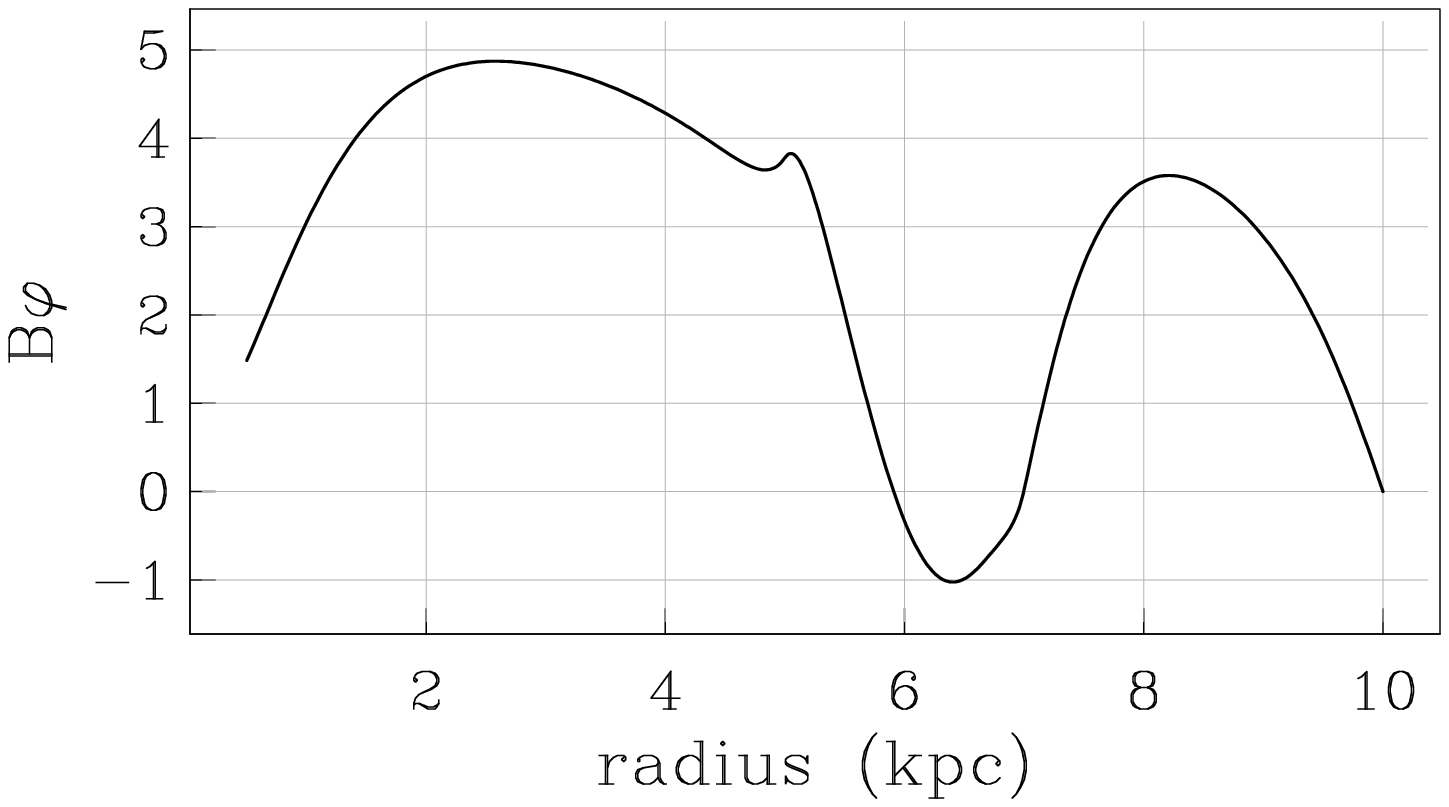}

\end{center}
\caption{The radial distribution of $B_\phi$ in the disc plane (a) immediately
before the counter-rotation is turned on; (b), (c), (d), (e) at times
$0.9, 1.8, 2.7, 3.6$ Gyr afterwards. 
}
\label{Bevol}
\end{figure}

Further models with $r_i=0.5 R_{\rm gal}, r_o=0.7 R_{\rm gal}$ with a) no counter-rotation,
$\alpha\propto \Omega, B_{\rm eq1}=0.1$, gave a strictly quadrupolar
$P=+1$; b) counter-rotation, $\alpha \propto |\Omega|$, $B_{\rm eq1}=0.1 B^*$,
resulted in $P=+0.98$.

In the counter-rotating case and $\alpha\propto \Omega$, as the dynamo numbers are reduced,
the contribution of the odd parity component also becomes smaller. For example,
with $R_\alpha=2, R_\omega=25$ in the steady state, the final parity is $P\approx 0.12$,
with $R_\alpha=1, R_\omega=25$ we get $P\approx 0.47$, and $R_\alpha=1, R_\omega=20$ also gives a mixed parity final state. 
In the marginally
supercritical case $R_\alpha=1, R_\omega=15$, the steady field is purely
quadrupolar, $P=+1$; the field in the outer ring decreases in relative strength with reduction in dynamo numbers.

In summary,  we can deduce that mixed parity fields  (in which the
azimuthal field has no symmetry with respect to the central plane) are a robust feature of
the model with counter-rotation and $\alpha\propto \Omega$.
With $\alpha\propto |\Omega|$ the effect is much reduced.
The region with $\alpha <  0$ seems crucial to the effect, counter-rotation
with $\alpha>0$ everywhere produces only a small effect.
The computations described in this Section started from a weak seed field
of mixed parity distributed throughout the disc. 
With our standard $\eta_D=10^{26}$ cm$^2$ s$^{-1}$, the
eventual steady state can take times in excess of $10^{10}$ years to become
established for the smaller dynamo numbers,
and departure from very near quadrupolar symmetry can then take a 
substantial fraction of this time. Timescales can be substantially reduced by different choices of the field present at the onset of counter-rotation.
We emphasize again that these estimated
timescales are sensitive to the value of $\eta_D$. Prolonged
timescales are plausibly associated with the diffusion of field across the gap between
the inner disc and the counter-rotating ring.

\subsection{Timescales and evolution}
\label{timescale}
In Sect.~\ref{basic2D} we have presented the saturated steady magnetic configuration
for a particular model. The results presented using the no-$z$ approximation
suggest that evolution to the final configuration may be quite slow -- this
feature is associated with the diffusion of field across the gap as the
final configuration is determined. 
If we start with an arbitrary seed field then, depending
on its strength, evolution to the final saturated configuration can take
times longer than the estimated lifetimes of rings, depending
on their believed mechanism of formation. For example, with rings
formed by accretion Buta \& Combes (1997) quote times that may approach the
Hubble time, whereas for rings resulting from impacts the models
of Athanassoula et al. (1996) suggest lifetimes of  less than 1~Gyr.
We show in Fig.~\ref{evol_seed} the evolution of parity and energy for an example where the seed field is weak (magnitude much smaller than equipartition strength), 
and of arbitrary parity.
After an initial brief fall in parity, a purely quadrupolar configuration 
persists for about 4 Gyr. The nonlinear regime is then reached and the field 
moves rapidly to the asymptotic state shown in Fig.~\ref{2D2fig}b.
However significant large-scale fields can be present in the ring regions
after $3-4$ Gyrs if the seed field is stronger.
Thus we now can consider two evolutionary
scenarios: either the counter-rotating ring is present from time $t=0$ with a 
seed field of unknown strength and geometry, or a merger event occurs creating the ring after a standard
even parity field has been established.

To illustrate this, with a stronger seed field a significant large-scale field
of magnitude comparable with equipartition strength can 
be established in the ring after $2-3$ Gyr, although the dynamo is far
from saturation, and a mixed parity field does not appear for another $0.5-1$ Gyr, with a longer time until the asymptotic state of Fig.~\ref{2D2}b is attained.

On the other hand, if the ring forms after an encounter or merger event
involving one or two pre-magnetised galaxies with near saturated dynamos, 
then the effective seed field 
for subsequent dynamo action may be quite strong and organised.
(We do not consider here the possibility that such an event generates turbulence
and small-scale magnetic fields of approximately equipartition strength,
which would provide a seed for subsequent dynamo action -- see, e.g., Moss et al. (2013).)
To gain some insight into this scenario, we allowed a standard, even parity, steady
global field to become established in a model with no counter-rotation, and 
no gap, with reduced strength equipartion field. Counter-rotation was then 
introduced, with $B_{\rm eq1}=0.1$, i.e. the model of Sect.~\ref{basic2D}.
In Fig.~\ref{Bevol} we show the dependence of $B_\phi$ in the plane $z=0$ on 
radius at times $0, 0.9, 1.8, 2.7$ and $3.6$ Gyr after the introduction of counter-rotation.
A significant field is present in the ring ($r>7$ kpc) throughout.

We can conclude that mixed parity fields may be attainable within galactic
timescales, depending on the details and history of the model. 
Other unexplored possibilities include that we underestimated angular velocity
gradients near the gap, and thus underestimated the strength of the dynamo action 
in that region, or that the turbulent diffusivity is not constant
throughout the disc plane. 
We have shown that large-scale fields may be present in galactic rings,
and that there is also  the possibility that they may be of mixed parity.
With our current state of knowledge we cannot be more specific.

\section{Discussions and conclusions}

We  have demonstrated that dynamo drivers in ring galaxies, i.e. differential rotation and mirror-asymmetry of 
interstellar
turbulence, can be strong enough to excite magnetic fields both in the disc and in the rings.  Our expectation
is that such rings do contain large-scale magnetic field (probably slightly weaker than in the inner discs) that
give rises to polarized radio emission.

We have found  that the current concepts of galactic dynamo action tell us quite counter-intuitively that the strong angular velocity gradients associated
with
counter-rotation of galactic rings with respect to  that of the disc does not lead to a substantial increase of 
dynamo action
and magnetic field strength near the ring. In other words, observations of 
magnetic field in galaxies with counter-rotating rings have the potential
to provide a
strong test for current galactic dynamo concepts.

 Inter alia, this investigation demonstrates the limits of the no-$z$
 model. It is known to be satisfactory (and very useful) for studying fields
that are believed with a high degree of confidence to have strictly even
parity with respect to the galactic equator, but is clearly unreliable
where this symmetry cannot be guaranteed, as in the models of Sect.~\ref{2D} above, where the stable fields are of mixed parity.

A potentially important feature of thin rings is the possibility
of generating non-axisymmetric fields.  In our model, such fields can 
more readily be generated
if magnetic field propagates freely through the outer boundary of the ring.
However dynamo numbers have to be chosen quite selectively to obtain nonaxisymmetric fields, and the result suggests that axisymmetric large-scale fields 
are the normal case.
Potentially more importantly, a magnetic configuration that is asymmetric in respect to the 
central galactic plane (a mixed parity solution) can arise near to the gap 
between the disc and the counter-rotating ring.

We note that our paper does not cover all dynamo related problems that arise in
the context of ring galaxies studies. 
In particular, it looks natural to investigate possible effects of dynamo action in resonant rings as well as 
the dynamo driven evolution of a strong (i.e. comparable with the 
equipartition field strength) and maybe small-scale seed field. 
A further possibility is the effects of ongoing injections
of small-scale fields of approximately equipartition strength  generated
by turbulent dynamo action, as in Moss et al. (2013).
Such extensions of the topic should however be  addressed separately.

We further note that in order to reach steady-state magnetic configurations we
sometimes run our code for 
times which are nominally several times longer than the anticipated 
lifetimes of the rings. However, the eventual steady state configurations
may not be so significant -- see e.g. Fig.~\ref{Bevol}.  
Timescales for our models, and their prior evolution, are quite uncertain. We have presented a scenario summarized in Fig.~\ref{Bevol} in which fields  can be present in galactic
rings more-or-less throughout the lifetime of the rings.
Whether the rings live long enough for global mixed parity configurations to
become established is more uncertain, although Buta \& Combes (1996)
do suggest that ring lifetimes may be quite extended. 
Observations of such fields 
would anyway be challenging.
We reiterate  that the timescale for 
magnetic field evolution depends on the value of the
turbulent magnetic diffusivity which is not a directly observable 
quantity and theoretical estimates contain substantial uncertainties. 
We have taken a standard figure ($10^{26}$ cm$^2$ s$^{-1}$),  
but need  to keep in mind that we may underestimate the
speed of magnetic field transport in ring galaxies. 
On the other hand, our results demonstrate that we can expect various long-lived
magnetic transients in ring galaxies,
 and contemporary magnetic configurations there can be influenced
by features of seed magnetic fields (or the field present after the ring forms).  

Chy\.zy \& Buta (2008) found  a coherent magnetic spiral structure crossing the inner pseudo-ring of NGC~4736. It is 
not exactly comparable with the outer rings discussed here,  
but it  appears to be an 
interesting system. 

We obtained an unexpected result for dynamo models with counter-rotating disc and ring. In spite of a very large 
shear between disc and ring, the resulting dynamo action is quite moderate. 
Even with our definition
of dynamo number $D$ the excitation threshold is found numerically to be
around  $D=13$, instead of $D=7.5$ for spiral 
galaxies. Thus the counter-rotating configuration inhibits the efficiency of dynamo action. The eigenfunction near marginal excitation  is concentrated
in the inner regions, and so this increase in marginal dynamo number 
is plausibly associated with its reduced radial scale,
and an opposition between the inner and outer dynamo drivers.      

In general, we can conclude that magnetic field studies in ring galaxies  have the potential to provide useful information 
concerning the nature and evolution of such galaxies.

\begin{acknowledgements}
EM acknowledges support from the Dynasty Foundation and RFBR under grant 16-32-00056. DS acknowledges support from RFBR under grant 15-02-01407. RB
is grateful for support from DFG Research Unit FOR1254. Analysis of the observational data on ring galaxies is supported by the RNF grant N 14-22-0041, 
and dynamo modelling of the counter-rotating ring by RNF grant 16-17-10097.
The authors thank Marita Krause, MPIfR's internal referee, for helpful comments,
and the external referee, Andrew Fletcher for suggesting a number of
improvements to the text.
\end{acknowledgements}

\end{document}